\documentclass[amsmath,amssymb,aps,showpacs,twocolumn,pra,prd,superscriptaddress,longbibliography,notitlepage]{revtex4-2}
\usepackage[dvips]{graphicx}
\usepackage{amsmath,amssymb,amsthm,mathrsfs,amsfonts,dsfont}
\usepackage{subfigure, epsfig}
\usepackage{braket}
\usepackage{bm}
\usepackage{bbm}
\usepackage{enumerate}
\usepackage{color}
\usepackage{comment}
\usepackage{diagbox}
\usepackage{youngtab}
\usepackage{lineno}
\usepackage{upgreek}
\usepackage[colorlinks=true,citecolor=blue]{hyperref}

\graphicspath{{./figure/}}

\newcommand{\comments}[1]{}

\begin{document}
% \linenumbers

\title{Quantum defects of $n$F$_J$ levels of Cs Rydberg atoms}

\author{Jingxu Bai}
\altaffiliation{These authors contributed equally to this work}
\affiliation{State Key Laboratory of Quantum Optics and Quantum Optics Devices, Institute of Laser Spectroscopy, Shanxi University, Taiyuan 030006, China}

\author{Yuechun Jiao}
\email{ycjiao@sxu.edu.cn}
\altaffiliation{These authors contributed equally to this work}
\affiliation{State Key Laboratory of Quantum Optics and Quantum Optics Devices, Institute of Laser Spectroscopy, Shanxi University, Taiyuan 030006, China}
\affiliation{Collaborative Innovation Center of Extreme Optics, Shanxi University, Taiyuan 030006, China}

\author{Rong Song}%
\affiliation{State Key Laboratory of Quantum Optics and Quantum Optics Devices, Institute of Laser Spectroscopy, Shanxi University, Taiyuan 030006, China}

\author{Jiabei Fan}%
\affiliation{State Key Laboratory of Quantum Optics and Quantum Optics Devices, Institute of Laser Spectroscopy, Shanxi University, Taiyuan 030006, China}

\author{Jianming Zhao}%
\email{zhaojm@sxu.edu.cn}
\affiliation{State Key Laboratory of Quantum Optics and Quantum Optics Devices, Institute of Laser Spectroscopy, Shanxi University, Taiyuan 030006, China}
\affiliation{Collaborative Innovation Center of Extreme Optics, Shanxi University, Taiyuan 030006, China}

\author{Suotang Jia}%
\affiliation{State Key Laboratory of Quantum Optics and Quantum Optics Devices, Institute of Laser Spectroscopy, Shanxi University, Taiyuan 030006, China}
\affiliation{Collaborative Innovation Center of Extreme Optics, Shanxi University, Taiyuan 030006, China}

\author{Georg Raithel}%
\email{graithel@umich.edu}
\affiliation{Department of Physics, University of Michigan, Ann Arbor, Michigan 48109-1120, USA}

\begin{abstract}
We present precise measurements of the quantum defects of cesium $n$F$_J$ Rydberg levels. We employ high-precision microwave spectroscopy of $(n+2)\mathrm{D}_{5/2}\rightarrow n\mathrm{F}_{5/2,7/2}$ transitions for $n=45$ to 50 in a cold-atom setup.  
Cold cesium $(n+2)$D$_{5/2}$ atoms, prepared via two-photon laser excitation, are probed by scanning weak microwave fields interacting with the atoms across the $n\mathrm{F}_{5/2,7/2}$ resonances. Transition spectra are acquired using state-selective electric-field ionization and time-gated ion detection. Transition-frequency intervals are obtained by Lorentzian fits to the measured spectral lines, which have linewidths ranging between 70~kHz and 190~kHz, corresponding to about one to three times the Fourier limit. A comprehensive analysis of relevant line-shift uncertainties and line-broadening effects is conducted. We find quantum defect parameters
$\delta_{0}(\mathrm{F}_{5/2})=0.03341537(70)$ and 
$\delta_{2}(\mathrm{F}_{5/2})=-0.2014(16)$, as well as 
$\delta_{0}(\mathrm{F}_{7/2})=0.0335646(13)$ and 
$\delta_{2}(\mathrm{F}_{7/2})=-0.2052(29)$, for $J=5/2$ and $J=7/2$, respectively. Fine structure parameters
$A_{FS}$ and $B_{FS}$ for Cs $n{\rm{F}}_J$ are also obtained. Results are discussed in context with previous works, and the significance of the results is discussed.
\end{abstract}

 \date{\today}% It is always \today, today,

\maketitle

\section{Introduction}

Accurate values for energy levels and quantum defects of alkali atoms, including those for high-angular-momentum states, play an ever-increasing role in testing atomic-structure and quantum-defect theories, as well as in  applications such as atomic clocks~\cite{Bloom,Martin,Milner}, quantum optics~\cite{Zhao,Wilk,Isenhower}, and Rydberg-atom-based metrology~\cite{Carter,Sedlacek}. Further, accurate information on atomic levels, atomic interactions, AC shifts etc, covered in our work, is important in the design of quantum gates in neutral-atom quantum computing and quantum simulation~\cite{Scholl2021,Wu2021}.
High-precision microwave spectroscopy is well-suited for precise measurement of Rydberg transition frequencies due to long atom-field interaction times and low transition line-widths that can be realized in slowly expanding cold-atom clouds. In combination with effective static-field control to reduce line shifts and broadening, accurate and precise transition-frequency measurements allow for the extraction of atomic constants.
This includes the quantum defects of low- and high-angular-momentum Rydberg states, which are affected by short-range many-electron interactions in the ionic Rydberg-atom core as well as long-range dipolar and quadrupolar long-range interaction between valence electron and ionic core, respectively~\cite{gallagher1994}. For alkali Rydberg atoms in high-$\ell$ ($\ell \ge 3$) states, the quantum defect arises from the ionic-core polarizabilities and the fine structure. Precision measurements of high-$\ell$ quantum defects therefore allow one to extract core polarizabilities and fine-structure coupling constants, enabling comparison with and validation of advanced atomic-structure calculations~\cite{Safronova, Cheung, Kaur, Beloy,Derevianko}.

Measurements of the quantum defects and ionization energy of cesium date back as far as 1949~\cite{Kratz,McNally}. Energy levels of Cs have been determined using classical methods such as grating and interference spectroscopy~\cite{Kleiman}, by means of a high-dispersion Czerny-Turner spectrograph and a hollow-cathode discharge~\cite{Eriksson}, and high-resolution Fourier spectroscopy~\cite{Sansonetti}. These measurements were for low principal quantum numbers, $n$, and had state-of-the art accuracy at that time. In addition, Doppler-free laser spectroscopy~\cite{Lorenzen,Lorenzen1983,Lorenzen1984,Sullivan} has been used to study Cs energy levels. Bjorkholm and Liao~\cite{Bjorkholm,Liao} proposed a method for resonance-enhanced Doppler-free multiphoton excitation, which was experimentally applied by Sansonetti and Lorenzen to measure fine structures for the odd-parity states of Cs. Goy $et~al.$~\cite{Goy} have measured quantum defects for S, P, D, and F Rydberg levels of Cs ($23<n<45$) by high-resolution double-resonance spectroscopy with an accuracy of about one MHz. In 1987, Weber $et~al.$~\cite{Weber} extensively measured absolute energy levels of Cs $n\mathrm{S}_{1/2}$, $n\mathrm{P}_{1/2}$, $n\mathrm{D}_{5/2}$, $n\mathrm{F}_{5/2}$, and $n\mathrm{G}_{7/2}$ Rydberg states using non-resonant and resonance-enhanced Doppler-free two-photon spectroscopy and obtained quantum defects. There, the laser wavelengths were measured using high-precision Fabry-Perot cavities with an uncertainty of 0.0002~$\mathrm{cm}^{-1}$ at most levels. These data are being widely used as reference values to this day.

In the present work, we laser-excite $n\mathrm{D}_{5/2}$ Rydberg states of Cs using a two-photon excitation scheme, and perform high-precision microwave spectroscopy of the electric-dipole $(n+2)\mathrm{D}_{5/2}\rightarrow n\mathrm{F}_{5/2,7/2}$ transitions. We describe our experimental procedures in some detail. We then obtain microwave spectra with resonance line-widths ranging between 70 kHz and
190 kHz, and determine transition frequencies with uncertainties of a few kHz. Using the Rydberg-Ritz formula, we extract the quantum defect parameters $\delta_0$ and $\delta_2$ for both fine structure components
$n\mathrm{F}_{5/2,7/2}$ of Cs, allowing us to also extract the fine-structure coupling constant, $A_{FS}$ and  $B_{FS}$. Results are discussed in context with previous works.

\section{Experimental setup}

\begin{figure}[htbp]
\centering\includegraphics[width=0.45\textwidth]{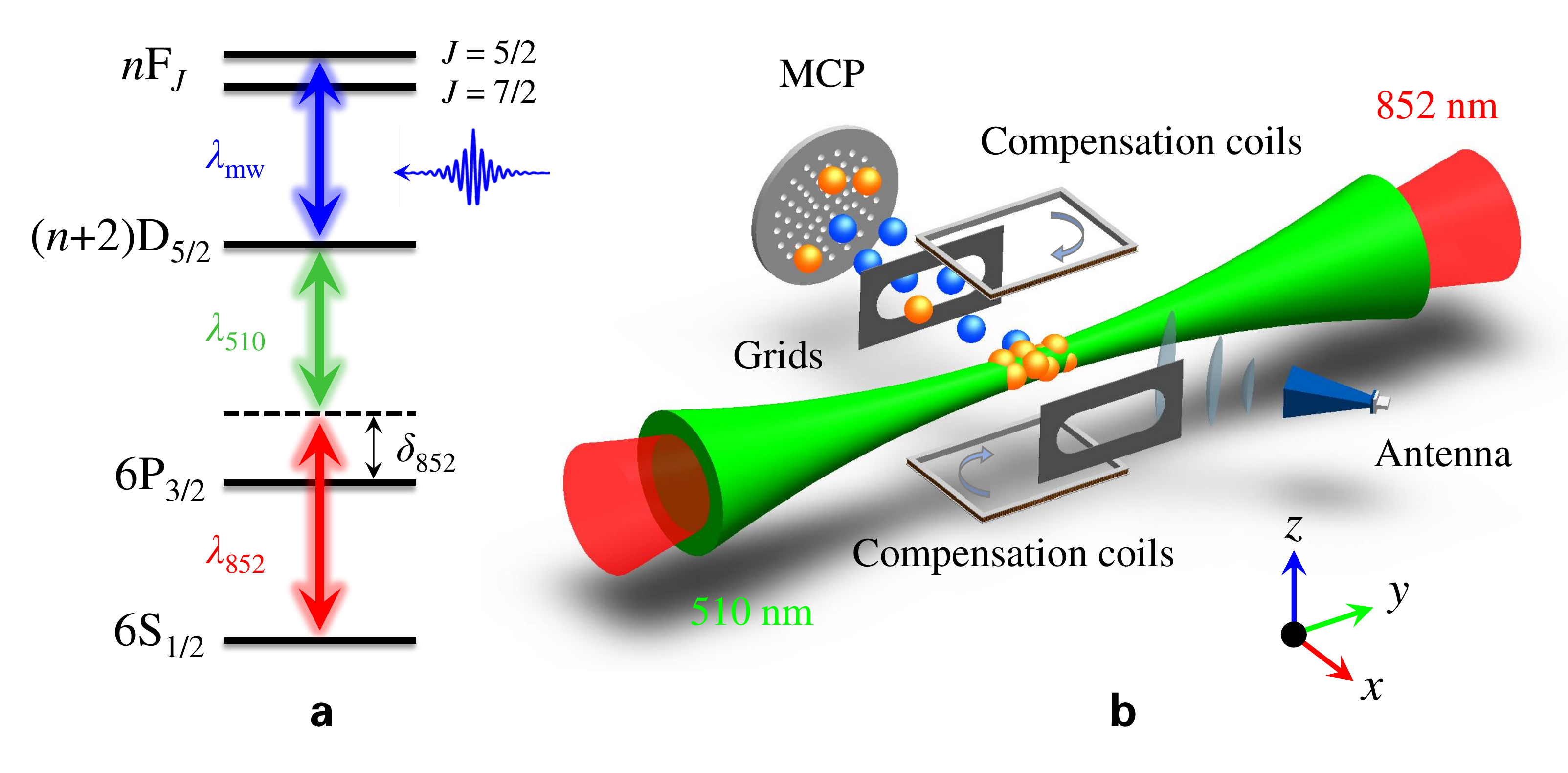}
\caption{(a) Level diagram and excitation scheme. 
The (n+2)$\mathrm{D}_{5/2}$ state is resonantly excited by two-photon excitation using $\lambda=852$-nm and 510-nm laser beams with an intermediate-state detuning of $\delta_{852}$ = +330~MHz. Microwaves $\lambda_{\mathrm{mw}}$ drive Rydberg transitions of the type $(n+2)\mathrm{D}_{5/2}\rightarrow n\mathrm{F}_{5/2,7/2}$. (b) Schematic of the experimental setup (not to scale). The excitation lasers are counter-propagated through the MOT center along the $y$ direction. The microwaves are introduced as shown. The $(n+2)$D$_{5/2}$ and microwave-excited $n$F$_{5/2}$ and $n$F$_{7/2}$ Rydberg atoms are counted by state-selective electric-field ionization and time-gated ion detection with an MCP detector. Three pairs of grids and compensation coils are placed on the $x$-, $y$- and $z$-axes to compensate stray electric and magnetic fields (only one set of each shown). Gold and blue balls represent laser-excited $(n+2)$D and microwave-coupled $n$F atoms.}\label{Fig1}
\end{figure}

In Fig.~\ref{Fig1}~(a) we display the level diagram used in our experiment. Atoms in $(n+2)$D$_{5/2}$ states ($n=45-50$) are populated using the displayed two-photon excitation scheme. The intermediate-state detuning of
$\delta_{852}$ = +330~MHz eliminates photon scattering and radiation pressure. The laser pulses have a duration of 500 ns. The density of the prepared Rydberg-atom sample is $\lesssim 4\times10^{6} \mathrm{cm}^{-3}$. Subsequent to laser preparation, a microwave pulse of  20-$\mu$s duration is applied to drive a Rydberg transition of the type $(n+2)\mathrm{D}_{5/2}\rightarrow n\mathrm{F}_{5/2,7/2}$, yielding narrow-linewidth microwave spectra.

Selected details of the experimental setup are shown in Fig.~\ref{Fig1}~(b). Cs ground-state atoms are laser-cooled and -trapped in a standard magneto-optical trap (MOT) with a temperature $\thicksim 100~\mathrm{\mu}$K and a peak density $\thicksim 10^{10}~\mathrm{cm}^{-3}$. After switching off the MOT and waiting for a delay time of 1~ms, we apply the 852-nm and 510-nm Rydberg-excitation laser pulses, which have respective Gaussian beam-waist
parameters of $w_0=750~\mu$m and 1000~$\mu$m.
Both lasers are external-cavity diode lasers from Toptica
that are locked to a high finesse Fabry-Perot cavity, resulting in laser linewidths of less than 100~kHz. The microwaves are generated by an analog signal generator (Keysight N5183B, frequency range 9~kHz to 40~GHz). The microwave output power and frequency scans are controlled with a Labview program. After turning off the microwave field, an electric-field ramp is applied to the grids on the $x$-axis for state-selective field ionization of the Rydberg atoms~\cite{gallagher1994}. Due to their different ionization limits, the laser-prepared $(n+2)$D and microwave-coupled $n$F Rydberg atoms result in ion signals at different arrival times on the microchannel plate (MCP) detector, allowing state-selective recording with a boxcar and a data acquisition card. Subsequent data analysis yields spectra as shown in Fig.~\ref{Fig2}.

Due to the large beam waists, the two-photon optical Rydberg-excitation Rabi frequency is fairy small; it is $\Omega$=$\Omega_{852} \Omega_{510}/(2 \delta_{852})$
% = $2\pi\times9.22$~kHz}
$\approx 2\pi\times9$~kHz.
Optical-pumping, saturation-broadening and radiation-pressure effects on the optical transitions are thus avoided. For our optical-pulse duration of 500~ns, the Rydberg-atom excitation probability per atom is $\sim 2 \times 10^{-4}$. The resultant Rydberg-atom density for a ground-state atom density in the MOT of $\lesssim 10^{10}~$cm$^{-3}$ is $\lesssim 4 \times 10^{6}~$cm$^{-3}$,
corresponding to a distance of $\sim 25~\mu$m between Rydberg atoms in the sample. As shown
in Sec.~\ref{subsec:densityshift}, this density is sufficiently low that transition shifts due to Rydberg-atom interactions are negligible, as required for high-precision microwave spectroscopy of Rydberg levels.

\section{Microwave spectroscopy}

\begin{figure}[htpb]
\centering\includegraphics[width=0.45\textwidth]{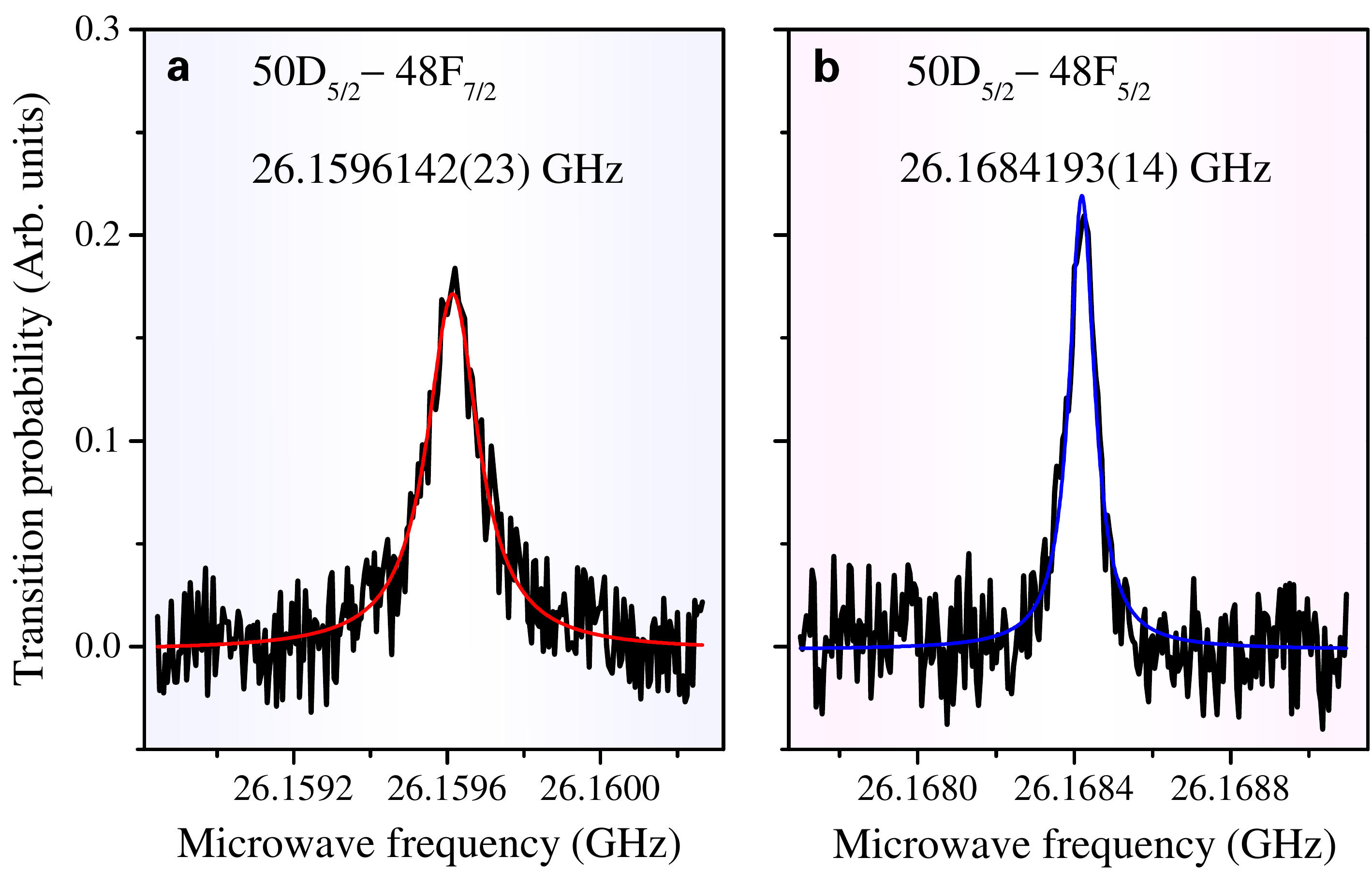}
\caption{ Measurements of the  microwave spectra for the $50\mathrm{D}_{5/2}$ to $48\mathrm{F}_{7/2}$ transition (a) and $48\mathrm{F}_{5/2}$ transition (b). To improve the signal-to-noise ratio, data are averaged over 25 or more repetitions of the experiment.
Black lines show the detected Rydberg transition probability signals. Curves in color are Lorentz fits to the spectra, yielding central frequencies of 26.1596142(23)~GHz and 26.1684193(14)~GHz, with statistical uncertainties, $\Delta \nu_{stat}$, in (). }\label{Fig2}
\end{figure}

In the experiment, we lock the frequencies of the two excitation lasers for resonant preparation  of Rydberg atoms in a $(n+2)$D$_{5/2}$ ($n$=45-50) state. The microwave frequency is scanned 
across the $(n+2)\mathrm{D}_{5/2}\rightarrow n\mathrm{F}_{5/2,7/2}$ transitions, driving the atoms between these states. High-precision microwave spectra of the $(n+2)\mathrm{D}_{5/2}\rightarrow n\mathrm{F}_{5/2,7/2}$ transitions are then obtained by state-selective field ionization and gated ion detection~\cite{gallagher1994}, and subsequent data analysis.

In Fig.~\ref{Fig2}, we show measured microwave spectra of the $50\mathrm{D}_{5/2}$ to $48\mathrm{F}_{7/2}$ (a) and $48\mathrm{F}_{5/2}$ (b) transitions. The duration of the microwave drive pulses is 20~$\mu$s. Lorentz fits (colored lines in Fig.~\ref{Fig2}) yield center frequencies of 26.1596142(23)~GHz and 26.1684193(14)~GHz for $50\mathrm{D}_{5/2}\rightarrow 48\mathrm{F}_{7/2}$ and $50\mathrm{D}_{5/2}\rightarrow 48\mathrm{F}_{5/2}$, respectively, with statistical uncertainties from the fits, $\Delta \nu_{stat}$, as indicated. 
%marked with black dashed lines. 
Due to the inversion of the fine structure of the Cs $n\mathrm{F}$ Rydberg levels, the $48\mathrm{F}_{7/2}$ level energy is lower than that of $48\mathrm{F}_{5/2}$.
%this is due to the influence of high-$l$ ($l$ > 4) manifolds. No, the reason is the many-electron physics near the core.  
The Lorentz fits also yield linewidths of about 163~kHz and 78~kHz for the $48\mathrm{F}_{7/2}$ and $48\mathrm{F}_{5/2}$ lines, respectively,
which is a factor of 1.5 to 3 larger than the Fourier-limited linewidth of 50~kHz for our 20-$\mu$s microwave drive pulses. 
In the following, line broadening and systematic line shifts will be discussed.

\section{Systematic effects}
\label{sec:syst}

First-off, the microwave generator must be locked to a reference clock. Further, Rydberg atoms are sensitive to stray DC electric fields due to their large polarizabilities ($\sim n^7$), to background DC magnetic fields due to the anomalous Zeeman effect, to AC shifts caused by the microwave drive field, and to interactions with other Rydberg atoms.
In order to obtain 
%[need to explain precision vs accuracy] 
measurements accurate to within a few kHz from the absolute line positions, required to determine the quantum defects of the $n$F$_{J}$ levels, and of the frequency intervals,
required to extract the fine structure parameters, systematic shifts due to all of these effects must be carefully considered.
For instance, it is imperative to compensate stray electric and magnetic fields at the MOT center using Stark and Zeeman microwave spectroscopy. We employ three independent pairs of electric-field grids and Helmholtz coils [not all shown in Fig.~\ref{Fig1}~(b)] to compensate the stray electric and magnetic fields to less than 2~mV/m and 5~mG, respectively.

\subsection{Signal generator frequency}

To obtain accurate transition-frequency readings, we use an external atomic clock (SRS FS725) as a reference to lock the crystal oscillator of the microwave generator. The clock's relative uncertainty is $\pm~5\times10^{-11}$, which results in a frequency deviation of less than 10~Hz~\cite{Moore} for the Rydberg transitions of interest. Hence, systematic shifts due to signal-generator frequency uncertainty are negligible. 

\subsection{Background DC magnetic fields}

While we carefully zero the magnetic field, an inhomogeneity of $\lesssim 5$~mG is sufficient cause several tens of kHz of broadening of the $(n+2)\mathrm{D}_{5/2} \to n \mathrm{F}_{J}$ lines. We have confirmed this by simulations of Zeeman spectra, which in the relevant magnetic-field range are in the linear (low-$B$) regime of the $(n+2)$D$_{5/2}$- and the
$n$F Rydberg-state fine structures, the Paschen-Back regime of the   
$n$F$_{J}$ hyperfine structure, and the transition regime of the 
$(n+2)$D$_{5/2}$ hyperfine structure. Assuming an unpolarized atomic sample and linear field polarizations, the Zeeman broadening is symmetric, consistent with the shapes of both spectral lines in Fig.~\ref{Fig2}. The unresolved hyperfine structure of the 
$(n+2)$D$_{5/2}$-levels, also included in the simulations, may add about $10$~kHz to the measured linewidths.

In detail, both experimental data and simulations of Zeeman spectra as a function of magnetic compensation fields $B_x$, $B_y$ and $B_z$ show symmetric and linear Zeeman splittings, as well as line-broadening effects that are also symmetric about the line centers. We have eliminated the Zeeman splittings and minimized residual magnetic-field-induced line broadening by adjustment of the compensation fields $B_x$, $B_y$ and $B_z$. The remnant magnetic field is estimated to be $\lesssim 5$~mG.
The symmetry of the Zeeman spectra about the line centers mostly results from the absence of circular polarization in the atom-field drives. The observed symmetry may be aided by the facts that the atomic sample is unpolarized, and that the remnant sub-5~mG fields are likely of quadrupolar character, without a preferred direction within the atomic sample. Based on the symmetry of any residual magnetic-field-induced broadening, as well as the small ($\lesssim 5$~mG) magnitude of any remnant magnetic fields, we assert that systematic magnetic shifts of the reported frequencies of $(n+2)\mathrm{D}_{5/2}\rightarrow n\mathrm{F}_{5/2,7/2}$ transitions are $\lesssim 1~$kHz. For the Zeeman-shift uncertainty we therefore use 1~kHz. We have verified the DC magnetic-field compensation every day throughout the data-taking.

\subsection{Background DC electric fields}

Stark effects typically cause asymmetric line broadening and, more importantly, a net line shift. Hence, special attention must be paid to compensating Stark effects. The electric field compensation process is described in~\cite{Bai}. Here we provide relevant details.

\begin{figure}[htpb]
\centering\includegraphics[width=0.45\textwidth]{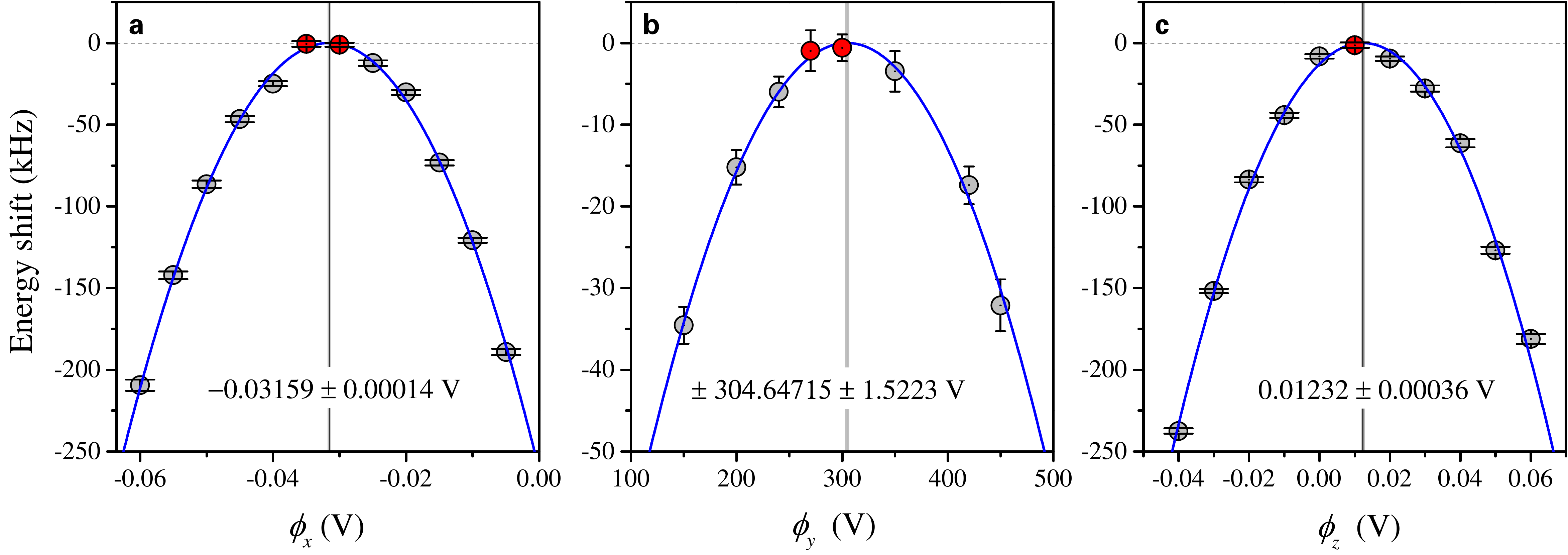}
\caption{Measurements of frequency shifts of the $50\mathrm{D}_{5/2} \to 48\mathrm{F}_{5/2}$ transition as a function of voltages, $\phi$, applied to electric-field compensation electrodes along the (a) $x$-, (b) $y$- and (c) $z$-direction, respectively. In (a) and (c) we apply the compensation voltage to one electrode, with the matching other connected to or referenced to ground [see Fig.~\ref{Fig1}~(b)]. In (b), we apply the voltage shown to both $y-$electrodes with opposite polarity, with the large overall magnitude of the compensation voltage being due to the large distance and efficient shielding of the $y-$electrodes from the location of the atoms. The data show the expected quadratic Stark shift behavior, which is in good agreement with simulations. The five red data points are used to estimate the systematic DC Stark shift of the transition frequencies reported in our work.}\label{Fig3}
\end{figure}

Fig.~\ref{Fig3} shows the peak shifts of microwave spectra as shown in Fig.~\ref{Fig2}, for the $50\mathrm{D}_{5/2} \to 48\mathrm{F}_{5/2}$ transition, as a function of voltages applied to the compensation electrodes along the $x-$, $y-$ and $z-$ directions. Zero shift is marked with a horizontal dashed line that shows the resonant frequency of the $50\mathrm{D}_{5/2} \to 48\mathrm{F}_{5/2}$ transition at zero field. Quadratic fits (blue curves) yield the voltages that must be applied to the respective electrodes for optimal DC electric-field compensation (vertical black lines; voltage values with fit uncertainties provided in the panels). 
The systematic uncertainty due to DC Stark shifts is estimated by computation of the RMS spread of the frequency shifts of the five data points closest to the apex of the three fit functions (red data points). The standard error of the mean of the average of these five points is 0.62~kHz, and the net uncertainty that follows from the uncertainties of the individual data points is 0.7~kHz. As conservative estimate of $\Delta \nu_{EDC}$ for the systematic DC Stark shift we therefore use $\Delta \nu_{EDC} = 1~$kHz. It should be noted that we have verified the DC stray-electric-field compensation on a daily basis in order to maintain optimal compensation throughout the entire data-taking sequence. Using the known polarizability of the transition frequency of $\Delta \alpha \approx 600~$kHz/(V/m)$^2$ (averaged over all values of the magnetic quantum number $m_j$) and noting that the Stark shift of the transition is $- \alpha E_{DC}^2/ 2$, a systematic DC Stark shift of 1~kHz translates into a DC electric-field uncertainty of about 2~mV/m at the location of the atoms.

\subsection{AC shifts}

\begin{figure}[htbp]
\centering\includegraphics[width=0.45\textwidth]{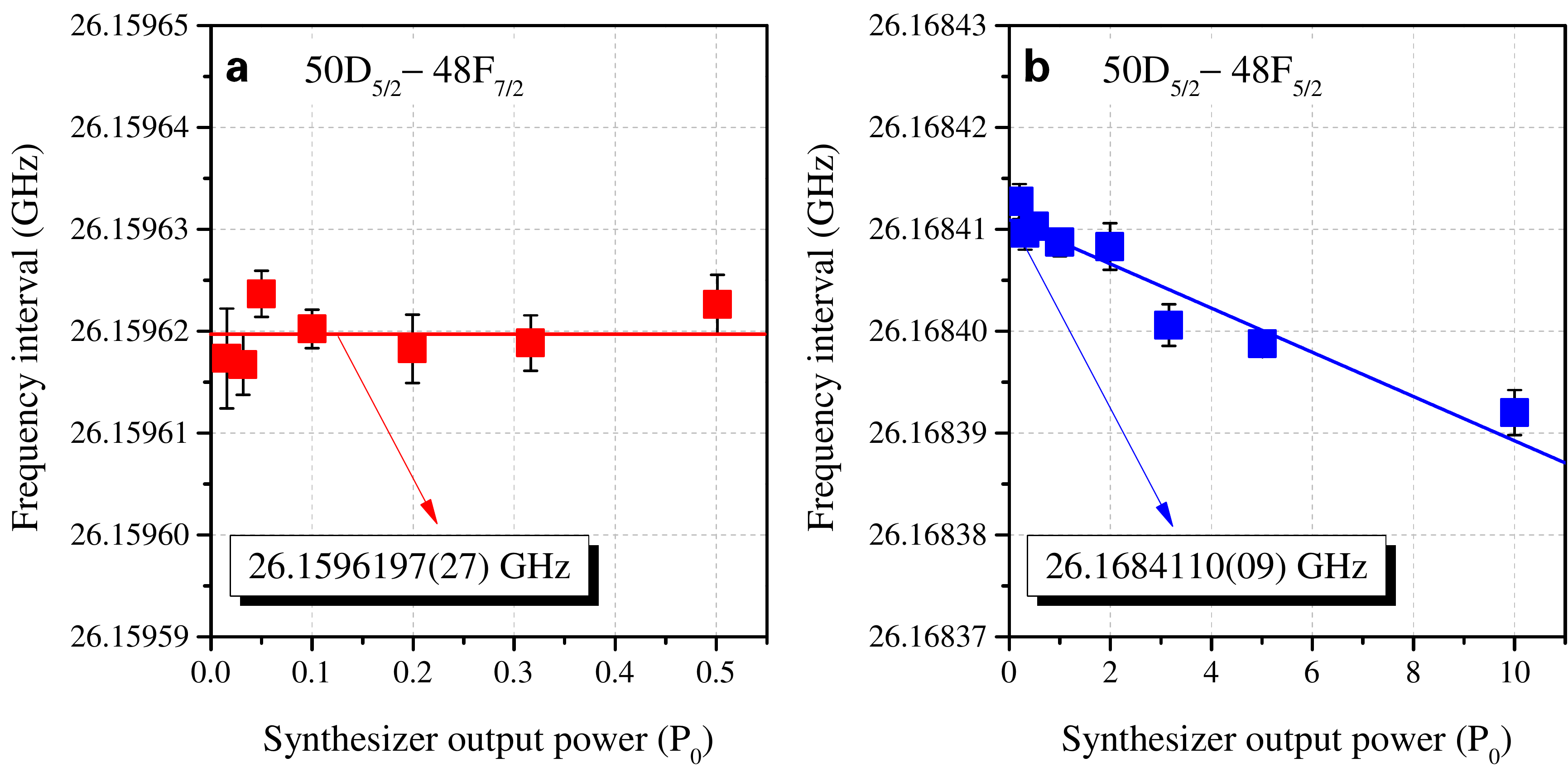}
\caption{ Frequencies of the $50\mathrm{D}_{5/2} \to  48\mathrm{F}_{7/2}$ (a) and $50\mathrm{D}_{5/2} \to  48\mathrm{F}_{5/2}$ (b) transitions vs synthesizer output power in units of $P_0 = -50$~dBm. Line centers and their uncertainties are obtained from Lorentzian fits to the microwave spectra. The measurement quantifies the AC shift on both transitions.}\label{Fig4}
\end{figure}

We next evaluate systematic shifts due to the AC Stark effect. The microwave intensity at the location of the atoms is varied by changing the synthesizer output power. In Figs.~\ref{Fig4}~(a) and~(b), we display the measured frequency interval as a function of the microwave output power. For the measured frequency interval of the $50\mathrm{D}_{5/2} \to
48\mathrm{F}_{7/2}$ transitions in Fig.~\ref{Fig4}~(a) and for microwave power less than 0.5~$P_0$, with reference level $P_0 = -50$~dBm on the signal generator, we find that the transition has no observable AC shift. The statistical variation of the line position over this microwave power range is less than $\approx 2$~kHz. Here, we may take multiple measurements within this power range and average the results to improve statistics. 
The data point for the transition frequency in Fig.~\ref{Fig4}~(a) is included in Table.~\ref{table1}.

For the $50\mathrm{D}_{5/2} \to 
48\mathrm{F}_{5/2}$ transition in Fig.~\ref{Fig4}~(b) we observe that the microwave power needed to 
drive the transition is on the order of a factor of twenty larger than for the $50\mathrm{D}_{5/2} \to 
48\mathrm{F}_{7/2}$ transition. This observation is in line with our computed microwave Rabi frequencies, which are about a factor of five lower for the $50\mathrm{D}_{5/2} \to  48\mathrm{F}_{5/2}$ than for the
$50\mathrm{D}_{5/2} \to 
48\mathrm{F}_{7/2}$ transition (averaged over $m_j$). Hence, we expect that the $50\mathrm{D}_{5/2} \to  48\mathrm{F}_{5/2}$ transition requires about a factor of 25 more in power than the $50\mathrm{D}_{5/2} \to  48\mathrm{F}_{7/2}$ transition to achieve spectral lines of similar height (as observed in the experiment).   
In addition, the $50\mathrm{D}_{5/2} \to  48\mathrm{F}_{5/2}$ transition has a significant AC shift, which is due to the higher microwave power required to drive that transition, and due to the 
large matrix element for the perturbing $50\mathrm{D}_{5/2} \to  48\mathrm{F}_{7/2}$ transition.
Comparing observed and calculated AC shifts, one may use Fig.~\ref{Fig4}~(b) to estimate microwave electric fields and Rabi frequencies. The estimation shows that at a power of $1 P_0$ the microwave electric field at the location of the atoms is $\sim 15$~mV/m, and the Rabi frequency is on the order of 50~kHz. This estimate accords well with the fact that a microwave power of about $1 P_0$ is required to saturate the $50\mathrm{D}_{5/2} \to  48\mathrm{F}_{5/2}$ transition with our $20~\mu$m-long microwave pulses. The lowest microwave fields used to probe the $\sim 25$ times stronger $50\mathrm{D}_{5/2} \to  48\mathrm{F}_{7/2}$ transition [see Fig.~\ref{Fig4}~(a)] are about $\sim 1$~mV/m. Following these consistency checks, we apply a linear fit to the data in Fig.~\ref{Fig4}~(b) to determine the y-intercept as our best estimate for the zero-microwave-field transition frequency of the $50\mathrm{D}_{5/2} \to  48\mathrm{F}_{5/2}$ transition. In Fig.~\ref{Fig4}~(b) and in analogous data in Table~\ref{table1} we report zero-microwave-field transition frequencies and the uncertainties of the y-intercepts that result from the linear fits. Hence, the effects of AC shifts are included in the statistical fit uncertainty and do not need to be accounted for separately.

\subsection{Shifts due to Rydberg-atom interactions}
\label{subsec:densityshift}

\begin{figure}[htbp]
\centering\includegraphics[width=0.45\textwidth]{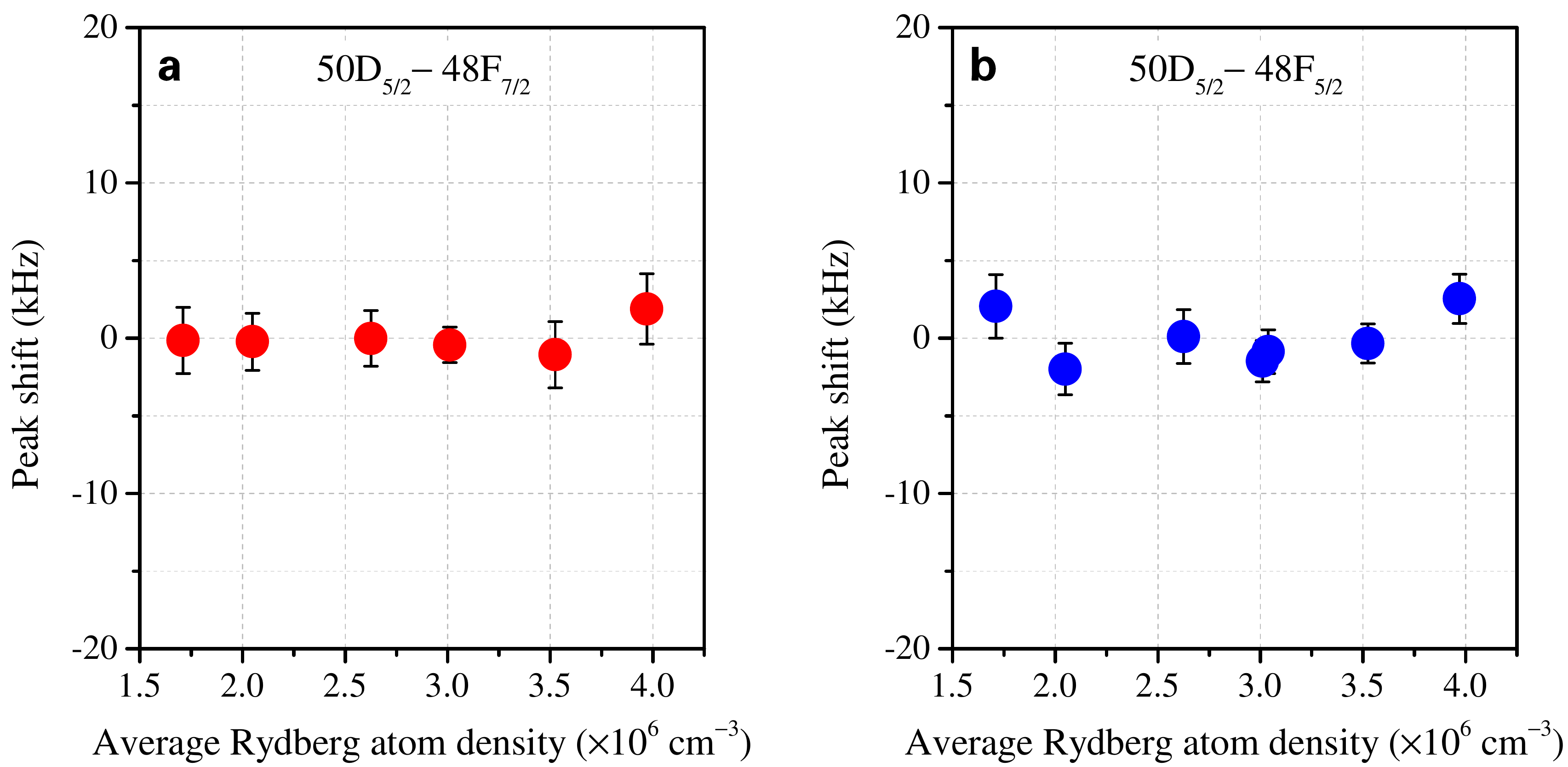}
\caption{
Line shifts of 
the $50\mathrm{D}_{5/2} \to  48\mathrm{F}_{7/2}$ (a) and $50\mathrm{D}_{5/2} \to  48\mathrm{F}_{5/2}$ (b) transitions vs estimated Rydberg-atom density. 
Line centers and their uncertainties are obtained from  Lorentzian fits to the microwave spectra. The measurements show that Rydberg-atom interactions have no measurable effect on the reported transition frequencies.}\label{Fig5}
\end{figure}

Finally we consider the effect of atomic interactions. Rydberg energy levels can be shifted due to an ubiquitous variety of dipolar and higher-order van-der-Waals interactions~\cite{Reinhard}, which scale as $n^{11}/R^6$ 
for the case of 2nd-order dipolar interaction ($R$ is the internuclear separation). However, in the present experiment the dominant interactions arise from the resonant electric-dipole coupling between atoms in $(n+2)\mathrm{D}_{5/2}$ and $n \mathrm{F}_J$ states, which become populated in the course of the optical and microwave excitations. Hence, the leading Rydberg-atom interaction scales as $n^4/R^3$. To explore potential transition-frequency shifts from these interactions, we vary the power of the 510-nm Rydberg-excitation laser in order to vary the Rydberg-atom density in the sample. As the optical two-photon excitation is far from saturation, this method presents an efficient means to vary the 
Rydberg-atom density. We then look for possible line shifts as a function of Rydberg-atom density.

Figs.~\ref{Fig5}~(a) and~(b) show the measured line shift as function of Rydberg-atom density for the $50\mathrm{D}_{5/2} \to
48\mathrm{F}_{7/2}$ and $50\mathrm{D}_{5/2} \to
48\mathrm{F}_{5/2}$ transitions, respectively. It can be seen that for 
estimated Rydberg-atom densities $\lesssim 4\times10^{6}~\mathrm{cm}^{-3}$, the density-induced line shift 
is less than the statistical variation of the data points, which is on the order of a few kHz in Fig.~\ref{Fig5}. 
Noting that average atom densities are somewhat below estimated peak densities, the internuclear separation between Rydberg atoms is estimated to be $\gtrsim 25 \mu$m at the highest densities in Fig.~\ref{Fig5}.

For a theoretical estimation of the density shifts, as a representative case we have computed binary molecular potentials for various combinations of Rydberg atoms in 50D$_{5/2}$ and 48F$_J$ states using methods developed in \cite{Han2018,Han2019}. It is found that the leading interaction between pairs of $50$D$_{5/2}$-state atoms is a dipolar van-der-Waals-type interaction, which scales as $C_6/R^6$, with a dispersion coefficient of $C_6 \sim 3.5$~GHz$\, \mu$m$^{6}$. The leading interaction between pairs of $48$F$_{J}$-state atoms also is a dipolar van-der-Waals-type interaction, with a dispersion coefficient of $C_6 \sim -20 $~GHz$\, \mu$m$^{6}$. The resultant level shifts at 30~$\mu$m internuclear separation amount to only tens of Hz and are negligible.
Atom pairs in a mix of 50D$_{5/2}$ and 48F$_J$-states interact much more strongly via resonant dipolar interactions, which scale as $C_3/R^3$. Here we find $|C_3| \sim 2.6$~GHz$\, \mu$m$^{3}$ for  50D$_{5/2}$ + 48F$_{7/2}$ pairs, and $|C_3| \sim 0.16$~GHz$\, \mu$m$^{3}$ for  50D$_{5/2}$ + 48F$_{5/2}$ pairs (values averaged over all molecular potentials for the allowed angular-momentum projections onto the internuclear axis). The $|C_3|$-coefficient for 50D$_{5/2}$ + 48F$_{7/2}$ is on the order of $(n^{* 2} e a_0)^2/(4 \pi \epsilon_0)$, the approximate maximum value for resonant dipolar coupling ($n^*$ is the effective quantum number). The large ratio between the $C_3$-coefficients for 50D$_{5/2}$ + 48F$_{7/2}$ versus 50D$_{5/2}$ +  48F$_{5/2}$ is due to angular matrix elements and is related to a likewise difference in Rabi-frequency-squares for the 50D$_{5/2} \to$ 48F$_{J}$ microwave transitions studied in our work. For atom pairs separated at $30~\mu$m, the {\sl{magnitudes}} of the line shifts are about 100~kHz for 50D$_{5/2} +$ 48F$_{7/2}$ and about 6~kHz for 50D$_{5/2} +$ 48F$_{5/2}$. The 
96 dipolar interaction potentials for 50D$_{5/2}$ + 48F$_{7/2}$ (and 72 for 50D$_{5/2}$ + 48F$_{5/2}$) are near-symmetric about the asymptotic energies (meaning there are as many attractive as there are repulsive potentials, and the magnitudes of positive and negative $C_3$-coefficients tend to be equal), and they are fairly evenly spread. For atom pairs separated at $30~\mu$m, the {\sl{average}} line shifts are only 26~Hz for 50D$_{5/2} +$ 48F$_{7/2}$ and 150~Hz for 50D$_{5/2} +$ 48F$_{5/2}$. Hence, the line shifts due to Rydberg-atom interactions are negligible, and the main effect of the interactions is a line broadening of the 50D$_{5/2} \to$ 48F$_{7/2}$ microwave transition, without causing significant shifts. The simulation results are in line with the experimental observations in Fig.~\ref{Fig5}, where no density-dependent line shift is measured for either transition. Hence, systematic shifts due to Rydberg-atom interactions are neglected in the present work. 

Summarizing our discussion of systematic shifts, in the following systematic uncertainties of 1~kHz are assumed for each of the DC electric and magnetic shifts. These are added in quadrature to the statistical uncertainties. All other systematic effects are either considerably smaller, or they are already covered by the statistical uncertainty of the measurements. The extra broadening of the $48\mathrm{F}_{7/2}$-line in Fig.~\ref{Fig2} and in analogous data for other close-by $n$-values is attributed to symmetric shifts due to Rydberg-atom interactions, which affects the $n\mathrm{F}_{7/2}$-lines about twenty times more than the $n\mathrm{F}_{5/2}$-lines. The dipolar Rydberg-atom interactions
have no measurable effect on the line centers, at our level of accuracy. It is noted that the differences in (symmetric) line broadening of the $48\mathrm{F}_{7/2}$- and $48\mathrm{F}_{5/2}$-lines shown in Fig.~\ref{Fig2}, as well as similar differences observed for other close-by $n$-values, cannot be attributed to differential effects of residual magnetic fields at the MOT center.

\section{Results and discussions}
\label{sec:results}

\subsection{Transition frequencies}
\label{subsec:freqs}

\begin{table}[htpb]
\caption{ Summary of measured transition frequencies and statistical uncertainties of $(n+2)\mathrm{D}_{5/2}\rightarrow n\mathrm{F}_{J}$ transitions.}\label{table1}
\begin{center}
	\renewcommand{\arraystretch}{1.7}
	\vspace{-2ex}
	\begin{tabular}{ccc}
		\hline\hline
		~$n$   &\qquad\qquad $\mathrm{F}_{5/2}$ &\qquad\qquad $\mathrm{F}_{7/2}$ \\
		\hline
		~45 &\qquad\qquad 31.793 121 7(13) &\qquad\qquad  31.782 448 6(30) \\
		~46 &\qquad\qquad 29.753 223 1(13) &\qquad\qquad  29.743 238 6(39) \\
		~47 &\qquad\qquad 27.884 148 5(11) &\qquad\qquad  27.874 789 8(18) \\
		~48 &\qquad\qquad 26.168 411 0(09) &\qquad\qquad  26.159 619 7(27) \\
		~49 &\qquad\qquad 24.590 605 6(37) &\qquad\qquad  24.582 325 2(16) \\
		~50 &\qquad\qquad 23.137 146 3(13) &\qquad\qquad  23.129 352 9(33) \\
		\hline\hline
	\end{tabular}
	\vspace{-3ex}
\end{center}
\end{table}

We have performed a series of microwave-spectroscopy measurements of $(n+2)\mathrm{D}_{5/2}\rightarrow n\mathrm{F}_{5/2,7/2}$ transitions for $n$ = 45-50. The extracted transition frequency intervals and their statistical uncertainties, $\Delta \nu_{stat}$ are listed in Table~\ref{table1}. It is seen there that the statistical uncertainties range between 1~kHz and 4~kHz. As discussed in Sec.~\ref{sec:syst}, AC-shift uncertainties are included in the statistical uncertainty, Rydberg-density shifts are neglected, and DC electric and Zeeman systematic uncertainties are 1~kHz each. Hence, the net uncertainties, $\Delta \nu$, used below, are given by 
\begin{equation}
\Delta \nu = \sqrt{ \Delta \nu_{stat}^2 + 2({\rm{kHz}}^2)} \quad,
\label{eq:unc}
\end{equation}
with $\Delta \nu_{stat}$ taken from Table~\ref{table1}.

\subsection{Quantum defects of $nF_J$-levels versus $n$}
\label{subsec:defects1}

The transition frequencies from an initial Rydberg state $(n, \ell, J)$ to a final state $(n', \ell', J')$
follow 
\begin{equation}
\nu_{n, \ell, J}^{n', \ell', J'} =  R_{\mathrm{Cs}} c \Big[\frac{1}{(n-\delta_{n, \ell, J})^{2}} - \frac{1}{(n^{'}-\delta_{n', \ell', J'})^{2}} \Big] \quad,
\label{eq:ryd1}
\end{equation}
with initial-state and final-state quantum defects $\delta_{n, \ell, J}$ and $\delta_{n', \ell', J'}$, respectively. There, $c=2.99792458\times10^{10}$~m/s is the speed of light and $R_{\mathrm{Cs}}= 109736.86273038(21)~\mathrm{cm}^{-1}$ is the Rydberg constant for Cs,
the uncertainty of which is dominated by the CODATA uncertainty of $R_\infty$, with the mass uncertainties of the Cs ion and the electron playing no significant role.

The quantum defects of the $nF_J$ states are then written as
\begin{equation}
\delta_{F}(n)=n-\Big( (n+2)^{*-2}_{D}-\frac{\nu}{cR_{Cs}} \Big)^{-1/2}=n-s^{-1/2} \quad,
\label{eq:ryd2}
\end{equation}
where $(n+2)_D^*$ is the effective quantum number of the $(n+2){\rm{D}}_{5/2}$ state, taken from~\cite{Weber}, and $s=(n+2)^{*-2}_{D}-\frac{\nu}{cR_{Cs}}$. 
From the measured frequency intervals in Table~\ref{table1}, $\nu$,  
we then obtain the quantum defects of $nF_J$ according to the Eq.~(\ref{eq:ryd2}).
The uncertainties of the measured quantum defects are
\begin{widetext}
\begin{equation}\label{eq:ryd3}
\Delta\delta_{F}(n)=\Big|\frac{1}{2}s^{-3/2} \Big| \sqrt{\big(\frac{\nu}{cR_{Cs}}\big)^{2}\big[(\frac{\Delta\nu}{\nu})^{2}+(\frac{\Delta R_{Cs}}{R_{Cs}})^{2} \big]+\big[
2(n+2)^{*-3}_{D} \Delta (n+2)_{D}^* \big]^{2}} \quad.
\end{equation}
\end{widetext}

The relative uncertainty of the Rydberg constant for Cs, $\Delta R_{Cs}/R_{Cs}$ = $1.9\times 10^{-12}$, contributes the least to $\Delta\delta_{F}$ (and has a negligible effect), followed by the frequency uncertainties of our measurements, $\Delta\nu / \nu$, from Table~\ref{table1} and Eq.~\ref{eq:unc}. The uncertainties $\Delta(n+2)^{*}_D$ of the quantum defects of the $(n+2)\mathrm{D}_{5/2}$ states from Table~V in~\cite{Weber} are the dominant source of uncertainty of our results for $\delta_{F}(n)$, listed in Table~\ref{table2}.

\begin{table}[htb]
\caption{Summary of quantum defects, $\delta_F(n)$, and their uncertainties from Eqs.~(\ref{eq:ryd2}) and~(\ref{eq:ryd3}), for $J=5/2$ and $J=7/2$.}\label{table2}
\begin{center}
	\renewcommand{\arraystretch}{1.7}
	\vspace{-2ex}
	\begin{tabular}{ccc}
		\hline\hline
		~$n$  &\qquad\qquad $\delta(\mathrm{F}_{5/2})$ &\qquad\qquad $\delta(\mathrm{F}_{7/2})$ \\
		\hline
		~45 &\qquad\qquad 0.03331591(65) &\qquad\qquad	0.03346340(65)	\\
		~46 &\qquad\qquad 0.03331988(65) &\qquad\qquad	0.03346726(65)	\\
		~47 &\qquad\qquad 0.03332401(65) &\qquad\qquad	0.03347137(65)  \\
		~48 &\qquad\qquad 0.03332788(65) &\qquad\qquad	0.03347533(65) \\
		~49 &\qquad\qquad 0.03333144(65) &\qquad\qquad	0.03347920(65) \\
		~50 &\qquad\qquad 0.03333469(65) &\qquad\qquad	0.03348245(65) \\
		\hline\hline
	\end{tabular}
	\vspace{-3ex}
\end{center}
\end{table}
We see that the measured quantum defects exhibit a significant increase with principal quantum number, $n$, both for $\mathrm{F}_{5/2}$ and $\mathrm{F}_{7/2}$. The uncertainties are all the same at the present level of precision because the uncertainties $\Delta (n+2)^{*}_D$ for the $(n+2)\mathrm{D}_{5/2}$ states from~\cite{Weber}, derived from the $A=\delta_0$- and $B=\delta_2$-values provided for D$_{5/2}$ in Table~V therein, are the dominant source of uncertainty for all measurements made.

\subsection{Quantum defect parameters $\delta_0$ and $\delta_2$ for $nF_J$}
\label{subsec:defects2}

The quantum defect can be written as~\cite{Li}
\begin{equation}
\delta(n)=\delta_{0}+ \frac{\delta_{2}}{(n-\delta_{0})^{2}} 
+ \frac{\delta_{4}}{(n-\delta_{0})^{4}}  + \ldots
\label{eq:qdef2}
\end{equation} 
where $\delta_0$ and $\delta_i$ ($i$=2,4,...) are the leading and higher-order 
quantum defect parameters. 
In general, when $n > 20$, we only consider the first two terms on the rhs in Eq.~(\ref{eq:qdef2}).

\begin{figure}[b]
\centering\includegraphics[width=0.45\textwidth]{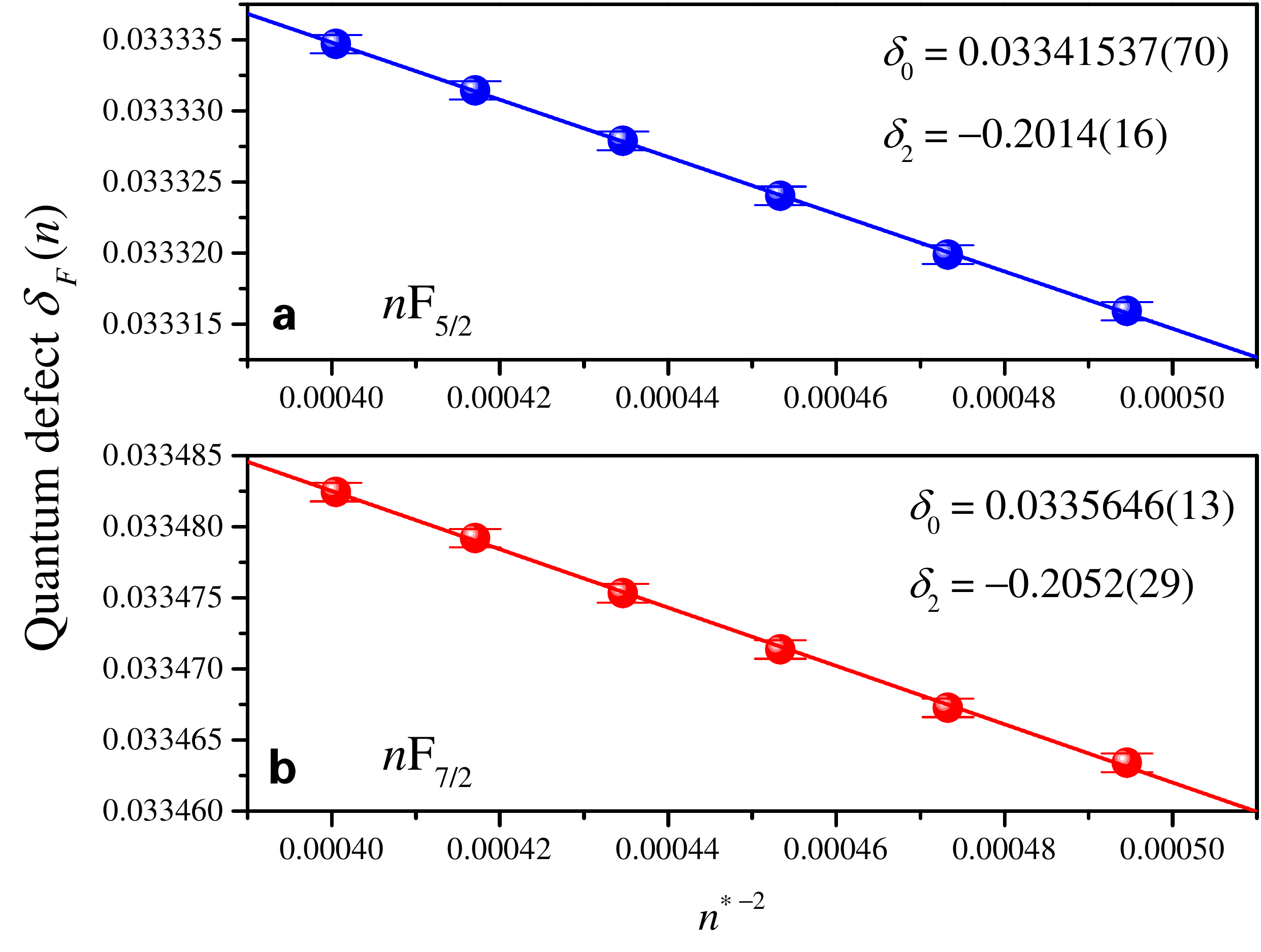}
\caption{Measured quantum defects $\delta_F(n)$ of $n$F$_J$ levels as a function of ${n^*}^{-2}$ for $n\mathrm{F}_{5/2}$ (a) and $n\mathrm{F}_{7/2}$ (b). The circles and uncertainty bars show the data from Table~\ref{table2}. 
	Blue ($n\mathrm{F}_{5/2}$) and red ($n\mathrm{F}_{7/2}$) solid lines represent fitting results 
	according to Eq.~(\ref{eq:qdef2}), yielding the  quantum defect parameters 
	$\delta_0(\mathrm{F}_{5/2})=0.03341537(70)$ and 
	$\delta_2(\mathrm{F}_{5/2})=-0.2014(16)$, and
	$\delta_0(\mathrm{F}_{7/2})=0.0335646(13)$ and 
	$\delta_2(\mathrm{F}_{7/2})=-0.2052(29)$, for $J=5/2$ and 7/2, respectively. 
}\label{Fig6}
\end{figure}

In Fig.~\ref{Fig6}, we plot the measured quantum defects $\delta_F$ for the $n$F$_J$ states from Table~\ref{table2} versus $n^{*-2}$. Fits according to Eq.~(\ref{eq:qdef2}) to the data then yield 
$\delta_0(\mathrm{F}_{5/2})=0.03341537(70)$ and 
$\delta_2(\mathrm{F}_{5/2})=-0.2014(16)$ for $nF_{5/2}$, and 
$\delta_0(\mathrm{F}_{7/2})=0.0335646(13)$ 
and $\delta_2(\mathrm{F}_{7/2})=-0.2052(29)$ for $nF_{7/2}$,
respectively. The quantum defect parameters $\delta_0$ and $\delta_2$ are listed in Table~\ref{table3}. To compare with previous works, we also list the quantum defects for $\mathrm{F}_J$ 
obtained in~\cite{Weber,Goy}. In~\cite{Weber}, the quantum defects were measured using nonresonant and resonantly enhanced Doppler-free two-photon spectroscopy of $n\mathrm{D}_{5/2}$, $n\mathrm{F}_{5/2}$, etc. There, the level energies were determined directly by measuring laser wavelengths with a high-precision Fabry-Perot interferometer, which had an uncertainty of 0.0002~$\mathrm{cm}^{-1}$ (6~MHz). Whereas, in~\cite{Goy} quantum defects were obtained by high-resolution double-resonance spectroscopy with $n$ ranging from 23 to 45 in the millimeter-wave domain, with an accuracy of about 1~MHz. 

In our work, we measure the quantum defects of the $n\mathrm{F}_J$-states by microwave spectroscopy with a narrow linewidth on the order of 100~kHz, from which accurate Rydberg-level frequency intervals are extracted with an uncertainty of $\sim$ 2~kHz. Since we have used $n$D$_{5/2}$ quantum defects from~\cite{Weber} as an input for our analysis, and since their uncertainties are the dominant source of uncertainty in our measurement, unsurprisingly for $n$F$_{5/2}$ our results in Table~\ref{table3}  agree within uncertainties with those from~\cite{Weber}; also, the uncertainties are comparable. A main deliverable of our work relies in a new set of quantum defect parameters $\delta_0$ and $\delta_2$ for $n$F$_{7/2}$,
for which reference~\cite{Weber} has no data. Our quantum defects are more precise than those from~\cite{Goy} by factors of 40 and 20, for $n$F$_{5/2}$ and $n$F$_{7/2}$, respectively. The values still agree within the uncertainties stated in~\cite{Goy}.

\begin{table}[htpb]
\caption{Quantum defect parameters $\delta_0$ and $\delta_2$ for $n\mathrm{F}_J$ levels from this work and several previous works~\cite{Goy,Weber}. }\label{table3}
\begin{center}
\renewcommand{\arraystretch}{1.8}
\vspace{-2ex}
\begin{tabular}{ccccc}
\hline\hline
~                     & $\delta_{0}$        &     $\delta_{2}$   &                    &              \\
\hline
~ $n\mathrm{F}_{5/2}$ & 0.03341537(70)  & $-0.2014(16)$  & This work     & ($45\leq n\leq 50$)  \\
~                     & 0.033392(30)   &  $-0.191(30)$   & ~\cite{Goy}   & ($23<n<45$)  \\
~                     & 0.03341424(96) &  $-0.198674$    & ~\cite{Weber} & ($~~6<n<65$) \\
\hline
~ $n\mathrm{F}_{7/2}$ &  0.0335646(13) &  $-0.2052(29)$  &  This work    & ($45\leq n\leq 50$)  \\
~                     &  0.033537(25)  &  $-0.191(20)$   &  ~\cite{Goy}  & ($23<n<45$)  \\
\hline\hline
\end{tabular}
\vspace{-1ex}
\end{center}
\end{table}

\subsection{Fine-structure intervals for $nF_J$}\label{subsec:fs}

\begin{figure}[htbp]
\centering\includegraphics[width=0.45\textwidth]{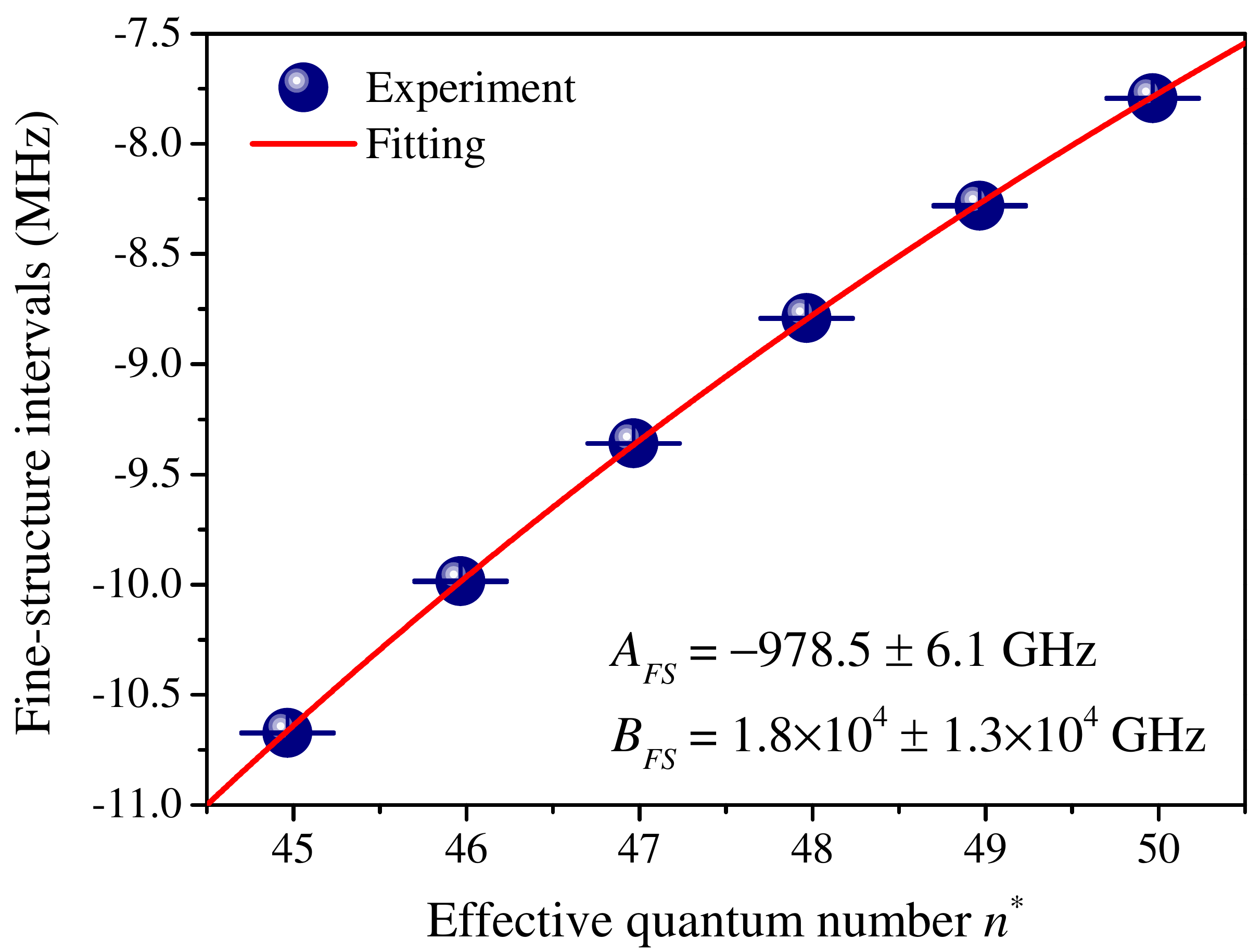}
\caption{Measured fine-structure frequency intervals between $n\mathrm{F}_{5/2}$ and $n\mathrm{F}_{7/2}$ as a function of effective quantum number $n^*$. The red solid line represents a 
	fit to the expression $A_{FS} \times n^{* -3} + B_{FS} \times n^{* -5}$, yielding fine structure parameters $A_{FS}=-978.5 \pm 6.1$~GHz and $B_{FS} = 1.8 \times 10^4 \pm 1.3 \times 10^4$~GHz, respectively. 
}\label{Fig7}
\end{figure}

The narrow spectral profiles allow us to clearly distinguish the fine structure of the $n$F Rydberg states for $J$=5/2 and 7/2, over our principal quantum number range of $n$ = 45 to 50 (limited by the frequency coverage of our microwave source).
From the frequency intervals $\nu$ for the $(n+2)\mathrm{D}_{5/2} \to n\mathrm{F}_{J}$ transitions in Table~\ref{table1}, we obtain the fine-structure splittings of the $n\mathrm{F}_J$-levels. 
In Fig.~\ref{Fig7}, we display the fine-structure splittings between $n\mathrm{F}_{5/2}$ and $n\mathrm{F}_{7/2}$ as a function of the effective quantum number $n^*$. Note that for Cs the $n$F fine structure is inverted, {\sl{i.e.}} $J=5/2$ is higher in energy than $J=7/2$.
The fine structure, attributed to the spin-orbit interaction of the Rydberg electron, is commonly described in terms of fine-structure coupling constants $A_{FS}$ and $B_{FS}$~\cite{gallagher1994}. The fine structure splitting is then 
given by $A_{FS} \times n^{* -3} + B_{FS} \times n^{* -5}$. Fitting the data in Fig.~\ref{Fig7}
with this expression, we find 
$A_{FS}=-978.5 \pm 6.1$~GHz and 
$B_{FS}$ = 1.8$\times 10^4 \pm 1.3 \times 10^4$~GHz.
These values are consistent with earlier results~\cite{Fredriksson1980} within our uncertainties
(we did not find an uncertainty in~\cite{Fredriksson1980}).

\section{Conclusion}

We have employed microwave spectroscopy and cold-atom methods to obtain the transition frequency intervals of cesium ($n$+2)$\mathrm{D}_{5/2} \to n\mathrm{F}_{5/2,7/2}$ transitions and have extracted quantum defects for the $n\mathrm{F}_{5/2,7/2}$ levels. A careful analysis of systematic uncertainties has been conducted. For $n \mathrm{F}_{5/2}$, our results and uncertainties agree well with data from~\cite{Weber}.  For $n \mathrm{F}_{7/2}$, our results are more precise than earlier data from~\cite{Goy} by a factor of about 20. 
The measured fine structure intervals have allowed us to also determine the $A_{FS}$ and $B_{FS}$ fine structure parameters for Cs $n{\rm{F}}$ levels; our results agree with previous data from~\cite{Fredriksson1980} to within our uncertainty.  

Our precise study of intrinsic properties of Rydberg atoms, such as quantum defects, is of significance 
for experimental work that relies on the availability of such data, as well as for theoretical research on the structure of complex atoms, where experimental data are often desired for a test of theoretical methods and results.
The uncertainties of our study were severely limited by the uncertainty of input data used for the quantum defects of our launch states, $n{\rm{D}}_{5/2}$. Based upon the frequency uncertainties $\Delta \nu$ we have already achieved in the present work, we expect that studies based upon alternate transition schemes, which are not reliant on previously-measured quantum defects, will allow us to improve our uncertainties for $n$F$_J$
quantum defects by a factor $\gtrsim 10$. Such methods may also enable studies of higher-$\ell$ quantum defects as well as refined studies of the hyperfine structure of Rydberg atoms. This may extend to the very small hyperfine structure of Cs $n \mathrm{D}$ Rydberg-states, which has not been previously studied (to our knowledge).

\section{Acknowledgments}
This work is supported by the National Natural Science Foundation of China (Grant Nos. 61835007, 12120101004, 62175136, 12241408); the Scientific Cooperation Exchanges Project of Shanxi province (Grant No. 202104041101015); the Changjiang Scholars and Innovative Research Team in Universities of the Ministry of Education of China (IRT 17R70); 1331 project of Shanxi province. GR acknowledges support by the University of Michigan.

\section{Disclosures}
The authors declare no conflicts of interest related to this article.

\section{Data Availability Statement}
Data underlying the results presented in this paper are not publicly available at this time but may be obtained from the authors upon reasonable request.

\section{AUTHOR CONTRIBUTIONS}
J.~Z and R.~G designed the study. J.~B and Y.~J collected and analyzed the data and wrote the original manuscript. R.~S, J.~F, and S.~J contributed to the manuscript revision. All authors provided review and comment on the subsequent versions of the manuscript. 

J.~B. and Y.~J. contributed equally to this work.

\bibliography{Fquantumdefect}

\providecommand{\noopsort}[1]{}\providecommand{\singleletter}[1]{#1}%
\begin{thebibliography}{36}%
\makeatletter
\providecommand \@ifxundefined [1]{%
 \@ifx{#1\undefined}
}%
\providecommand \@ifnum [1]{%
 \ifnum #1\expandafter \@firstoftwo
 \else \expandafter \@secondoftwo
 \fi
}%
\providecommand \@ifx [1]{%
 \ifx #1\expandafter \@firstoftwo
 \else \expandafter \@secondoftwo
 \fi
}%
\providecommand \natexlab [1]{#1}%
\providecommand \enquote  [1]{``#1''}%
\providecommand \bibnamefont  [1]{#1}%
\providecommand \bibfnamefont [1]{#1}%
\providecommand \citenamefont [1]{#1}%
\providecommand \href@noop [0]{\@secondoftwo}%
\providecommand \href [0]{\begingroup \@sanitize@url \@href}%
\providecommand \@href[1]{\@@startlink{#1}\@@href}%
\providecommand \@@href[1]{\endgroup#1\@@endlink}%
\providecommand \@sanitize@url [0]{\catcode `\\12\catcode `\$12\catcode
  `\&12\catcode `\#12\catcode `\^12\catcode `\_12\catcode `\%12\relax}%
\providecommand \@@startlink[1]{}%
\providecommand \@@endlink[0]{}%
\providecommand \url  [0]{\begingroup\@sanitize@url \@url }%
\providecommand \@url [1]{\endgroup\@href {#1}{\urlprefix }}%
\providecommand \urlprefix  [0]{URL }%
\providecommand \Eprint [0]{\href }%
\providecommand \doibase [0]{https://doi.org/}%
\providecommand \selectlanguage [0]{\@gobble}%
\providecommand \bibinfo  [0]{\@secondoftwo}%
\providecommand \bibfield  [0]{\@secondoftwo}%
\providecommand \translation [1]{[#1]}%
\providecommand \BibitemOpen [0]{}%
\providecommand \bibitemStop [0]{}%
\providecommand \bibitemNoStop [0]{.\EOS\space}%
\providecommand \EOS [0]{\spacefactor3000\relax}%
\providecommand \BibitemShut  [1]{\csname bibitem#1\endcsname}%
\let\auto@bib@innerbib\@empty
%</preamble>
\bibitem [{\citenamefont {Bloom}\ \emph {et~al.}(2014)\citenamefont {Bloom},
  \citenamefont {Nicholson}, \citenamefont {Williams}, \citenamefont
  {Campbell}, \citenamefont {Bishof}, \citenamefont {Zhang}, \citenamefont
  {Zhang}, \citenamefont {Bromley},\ and\ \citenamefont {Ye}}]{Bloom}%
  \BibitemOpen
  \bibfield  {author} {\bibinfo {author} {\bibfnamefont {B.}~\bibnamefont
  {Bloom}}, \bibinfo {author} {\bibfnamefont {T.}~\bibnamefont {Nicholson}},
  \bibinfo {author} {\bibfnamefont {J.}~\bibnamefont {Williams}}, \bibinfo
  {author} {\bibfnamefont {S.}~\bibnamefont {Campbell}}, \bibinfo {author}
  {\bibfnamefont {M.}~\bibnamefont {Bishof}}, \bibinfo {author} {\bibfnamefont
  {X.}~\bibnamefont {Zhang}}, \bibinfo {author} {\bibfnamefont
  {W.}~\bibnamefont {Zhang}}, \bibinfo {author} {\bibfnamefont
  {S.}~\bibnamefont {Bromley}},\ and\ \bibinfo {author} {\bibfnamefont
  {J.}~\bibnamefont {Ye}},\ }\bibfield  {title} {\bibinfo {title} {An optical
  lattice clock with accuracy and stability at the $10^{-18}$ level},\ }\href
  {https://doi.org/10.1038/nature12941} {\bibfield  {journal} {\bibinfo
  {journal} {Nature}\ }\textbf {\bibinfo {volume} {506}},\ \bibinfo {pages}
  {71} (\bibinfo {year} {2014})}\BibitemShut {NoStop}%
\bibitem [{\citenamefont {Martin}\ \emph {et~al.}(2018)\citenamefont {Martin},
  \citenamefont {Phelps}, \citenamefont {Lemke}, \citenamefont {Bigelow},
  \citenamefont {Stuhl}, \citenamefont {Wojcik}, \citenamefont {Holt},
  \citenamefont {Coddington}, \citenamefont {Bishop},\ and\ \citenamefont
  {Burke}}]{Martin}%
  \BibitemOpen
  \bibfield  {author} {\bibinfo {author} {\bibfnamefont {K.~W.}\ \bibnamefont
  {Martin}}, \bibinfo {author} {\bibfnamefont {G.}~\bibnamefont {Phelps}},
  \bibinfo {author} {\bibfnamefont {N.~D.}\ \bibnamefont {Lemke}}, \bibinfo
  {author} {\bibfnamefont {M.~S.}\ \bibnamefont {Bigelow}}, \bibinfo {author}
  {\bibfnamefont {B.}~\bibnamefont {Stuhl}}, \bibinfo {author} {\bibfnamefont
  {M.}~\bibnamefont {Wojcik}}, \bibinfo {author} {\bibfnamefont
  {M.}~\bibnamefont {Holt}}, \bibinfo {author} {\bibfnamefont {I.}~\bibnamefont
  {Coddington}}, \bibinfo {author} {\bibfnamefont {M.~W.}\ \bibnamefont
  {Bishop}},\ and\ \bibinfo {author} {\bibfnamefont {J.~H.}\ \bibnamefont
  {Burke}},\ }\bibfield  {title} {\bibinfo {title} {Compact optical atomic
  clock based on a two-photon transition in rubidium},\ }\href
  {https://doi.org/10.1103/PhysRevApplied.9.014019} {\bibfield  {journal}
  {\bibinfo  {journal} {Phys. Rev. Appl.}\ }\textbf {\bibinfo {volume} {9}},\
  \bibinfo {pages} {014019} (\bibinfo {year} {2018})}\BibitemShut {NoStop}%
\bibitem [{\citenamefont {Milner}\ \emph {et~al.}(2019)\citenamefont {Milner},
  \citenamefont {Robinson}, \citenamefont {Kennedy}, \citenamefont {Bothwell},
  \citenamefont {Kedar}, \citenamefont {Matei}, \citenamefont {Legero},
  \citenamefont {Sterr}, \citenamefont {Riehle}, \citenamefont {Leopardi},
  \citenamefont {Fortier}, \citenamefont {Sherman}, \citenamefont {Levine},
  \citenamefont {Yao}, \citenamefont {Ye},\ and\ \citenamefont
  {Oelker}}]{Milner}%
  \BibitemOpen
  \bibfield  {author} {\bibinfo {author} {\bibfnamefont {W.~R.}\ \bibnamefont
  {Milner}}, \bibinfo {author} {\bibfnamefont {J.~M.}\ \bibnamefont
  {Robinson}}, \bibinfo {author} {\bibfnamefont {C.~J.}\ \bibnamefont
  {Kennedy}}, \bibinfo {author} {\bibfnamefont {T.}~\bibnamefont {Bothwell}},
  \bibinfo {author} {\bibfnamefont {D.}~\bibnamefont {Kedar}}, \bibinfo
  {author} {\bibfnamefont {D.~G.}\ \bibnamefont {Matei}}, \bibinfo {author}
  {\bibfnamefont {T.}~\bibnamefont {Legero}}, \bibinfo {author} {\bibfnamefont
  {U.}~\bibnamefont {Sterr}}, \bibinfo {author} {\bibfnamefont
  {F.}~\bibnamefont {Riehle}}, \bibinfo {author} {\bibfnamefont
  {H.}~\bibnamefont {Leopardi}}, \bibinfo {author} {\bibfnamefont {T.~M.}\
  \bibnamefont {Fortier}}, \bibinfo {author} {\bibfnamefont {J.~A.}\
  \bibnamefont {Sherman}}, \bibinfo {author} {\bibfnamefont {J.}~\bibnamefont
  {Levine}}, \bibinfo {author} {\bibfnamefont {J.}~\bibnamefont {Yao}},
  \bibinfo {author} {\bibfnamefont {J.}~\bibnamefont {Ye}},\ and\ \bibinfo
  {author} {\bibfnamefont {E.}~\bibnamefont {Oelker}},\ }\bibfield  {title}
  {\bibinfo {title} {Demonstration of a timescale based on a stable optical
  carrier},\ }\href {https://doi.org/10.1103/PhysRevLett.123.173201} {\bibfield
   {journal} {\bibinfo  {journal} {Phys. Rev. Lett.}\ }\textbf {\bibinfo
  {volume} {123}},\ \bibinfo {pages} {173201} (\bibinfo {year}
  {2019})}\BibitemShut {NoStop}%
\bibitem [{\citenamefont {Zhao}\ \emph {et~al.}(2010)\citenamefont {Zhao},
  \citenamefont {M\"uller}, \citenamefont {Hammerer},\ and\ \citenamefont
  {Zoller}}]{Zhao}%
  \BibitemOpen
  \bibfield  {author} {\bibinfo {author} {\bibfnamefont {B.}~\bibnamefont
  {Zhao}}, \bibinfo {author} {\bibfnamefont {M.}~\bibnamefont {M\"uller}},
  \bibinfo {author} {\bibfnamefont {K.}~\bibnamefont {Hammerer}},\ and\
  \bibinfo {author} {\bibfnamefont {P.}~\bibnamefont {Zoller}},\ }\bibfield
  {title} {\bibinfo {title} {Efficient quantum repeater based on deterministic
  rydberg gates},\ }\href {https://doi.org/10.1103/PhysRevA.81.052329}
  {\bibfield  {journal} {\bibinfo  {journal} {Phys. Rev. A}\ }\textbf {\bibinfo
  {volume} {81}},\ \bibinfo {pages} {052329} (\bibinfo {year}
  {2010})}\BibitemShut {NoStop}%
\bibitem [{\citenamefont {Wilk}\ \emph {et~al.}(2010)\citenamefont {Wilk},
  \citenamefont {Ga\"etan}, \citenamefont {Evellin}, \citenamefont {Wolters},
  \citenamefont {Miroshnychenko}, \citenamefont {Grangier},\ and\ \citenamefont
  {Browaeys}}]{Wilk}%
  \BibitemOpen
  \bibfield  {author} {\bibinfo {author} {\bibfnamefont {T.}~\bibnamefont
  {Wilk}}, \bibinfo {author} {\bibfnamefont {A.}~\bibnamefont {Ga\"etan}},
  \bibinfo {author} {\bibfnamefont {C.}~\bibnamefont {Evellin}}, \bibinfo
  {author} {\bibfnamefont {J.}~\bibnamefont {Wolters}}, \bibinfo {author}
  {\bibfnamefont {Y.}~\bibnamefont {Miroshnychenko}}, \bibinfo {author}
  {\bibfnamefont {P.}~\bibnamefont {Grangier}},\ and\ \bibinfo {author}
  {\bibfnamefont {A.}~\bibnamefont {Browaeys}},\ }\bibfield  {title} {\bibinfo
  {title} {Entanglement of two individual neutral atoms using rydberg
  blockade},\ }\href {https://doi.org/10.1103/PhysRevLett.104.010502}
  {\bibfield  {journal} {\bibinfo  {journal} {Phys. Rev. Lett.}\ }\textbf
  {\bibinfo {volume} {104}},\ \bibinfo {pages} {010502} (\bibinfo {year}
  {2010})}\BibitemShut {NoStop}%
\bibitem [{\citenamefont {Isenhower}\ \emph {et~al.}(2010)\citenamefont
  {Isenhower}, \citenamefont {Urban}, \citenamefont {Zhang}, \citenamefont
  {Gill}, \citenamefont {Henage}, \citenamefont {Johnson}, \citenamefont
  {Walker},\ and\ \citenamefont {Saffman}}]{Isenhower}%
  \BibitemOpen
  \bibfield  {author} {\bibinfo {author} {\bibfnamefont {L.}~\bibnamefont
  {Isenhower}}, \bibinfo {author} {\bibfnamefont {E.}~\bibnamefont {Urban}},
  \bibinfo {author} {\bibfnamefont {X.~L.}\ \bibnamefont {Zhang}}, \bibinfo
  {author} {\bibfnamefont {A.~T.}\ \bibnamefont {Gill}}, \bibinfo {author}
  {\bibfnamefont {T.}~\bibnamefont {Henage}}, \bibinfo {author} {\bibfnamefont
  {T.~A.}\ \bibnamefont {Johnson}}, \bibinfo {author} {\bibfnamefont {T.~G.}\
  \bibnamefont {Walker}},\ and\ \bibinfo {author} {\bibfnamefont
  {M.}~\bibnamefont {Saffman}},\ }\bibfield  {title} {\bibinfo {title}
  {Demonstration of a neutral atom controlled-not quantum gate},\ }\href
  {https://doi.org/10.1103/PhysRevLett.104.010503} {\bibfield  {journal}
  {\bibinfo  {journal} {Phys. Rev. Lett.}\ }\textbf {\bibinfo {volume} {104}},\
  \bibinfo {pages} {010503} (\bibinfo {year} {2010})}\BibitemShut {NoStop}%
\bibitem [{\citenamefont {Carter}\ \emph {et~al.}(2012)\citenamefont {Carter},
  \citenamefont {Cherry},\ and\ \citenamefont {Martin}}]{Carter}%
  \BibitemOpen
  \bibfield  {author} {\bibinfo {author} {\bibfnamefont {J.~D.}\ \bibnamefont
  {Carter}}, \bibinfo {author} {\bibfnamefont {O.}~\bibnamefont {Cherry}},\
  and\ \bibinfo {author} {\bibfnamefont {J.~D.~D.}\ \bibnamefont {Martin}},\
  }\bibfield  {title} {\bibinfo {title} {Electric-field sensing near the
  surface microstructure of an atom chip using cold rydberg atoms},\ }\href
  {https://doi.org/10.1103/PhysRevA.86.053401} {\bibfield  {journal} {\bibinfo
  {journal} {Phys. Rev. A}\ }\textbf {\bibinfo {volume} {86}},\ \bibinfo
  {pages} {053401} (\bibinfo {year} {2012})}\BibitemShut {NoStop}%
\bibitem [{\citenamefont {Sedlacek}\ \emph {et~al.}(2012)\citenamefont
  {Sedlacek}, \citenamefont {Schwettmann}, \citenamefont {Kübler},
  \citenamefont {L\"{o}w}, \citenamefont {Pfau},\ and\ \citenamefont
  {Shaffer}}]{Sedlacek}%
  \BibitemOpen
  \bibfield  {author} {\bibinfo {author} {\bibfnamefont {J.}~\bibnamefont
  {Sedlacek}}, \bibinfo {author} {\bibfnamefont {A.}~\bibnamefont
  {Schwettmann}}, \bibinfo {author} {\bibfnamefont {H.}~\bibnamefont
  {Kübler}}, \bibinfo {author} {\bibfnamefont {R.}~\bibnamefont {L\"{o}w}},
  \bibinfo {author} {\bibfnamefont {T.}~\bibnamefont {Pfau}},\ and\ \bibinfo
  {author} {\bibfnamefont {J.~P.}\ \bibnamefont {Shaffer}},\ }\bibfield
  {title} {\bibinfo {title} {Microwave electrometry with rydberg atoms in a
  vapour cell using bright atomic resonances},\ }\href
  {https://doi.org/10.1038/nphys2423} {\bibfield  {journal} {\bibinfo
  {journal} {Nature Physics}\ }\textbf {\bibinfo {volume} {8}},\ \bibinfo
  {pages} {819} (\bibinfo {year} {2012})}\BibitemShut {NoStop}%
\bibitem [{\citenamefont {Scholl}\ \emph {et~al.}(2021)\citenamefont {Scholl},
  \citenamefont {Schuler}, \citenamefont {Williams}, \citenamefont
  {Eberharter}, \citenamefont {Barredo}, \citenamefont {Schymik}, \citenamefont
  {Lienhard}, \citenamefont {Henry}, \citenamefont {Lang}, \citenamefont
  {Lahaye}, \citenamefont {L{\"a}uchli},\ and\ \citenamefont
  {Browaeys}}]{Scholl2021}%
  \BibitemOpen
  \bibfield  {author} {\bibinfo {author} {\bibfnamefont {P.}~\bibnamefont
  {Scholl}}, \bibinfo {author} {\bibfnamefont {M.}~\bibnamefont {Schuler}},
  \bibinfo {author} {\bibfnamefont {H.~J.}\ \bibnamefont {Williams}}, \bibinfo
  {author} {\bibfnamefont {A.~A.}\ \bibnamefont {Eberharter}}, \bibinfo
  {author} {\bibfnamefont {D.}~\bibnamefont {Barredo}}, \bibinfo {author}
  {\bibfnamefont {K.-N.}\ \bibnamefont {Schymik}}, \bibinfo {author}
  {\bibfnamefont {V.}~\bibnamefont {Lienhard}}, \bibinfo {author}
  {\bibfnamefont {L.-P.}\ \bibnamefont {Henry}}, \bibinfo {author}
  {\bibfnamefont {T.~C.}\ \bibnamefont {Lang}}, \bibinfo {author}
  {\bibfnamefont {T.}~\bibnamefont {Lahaye}}, \bibinfo {author} {\bibfnamefont
  {A.~M.}\ \bibnamefont {L{\"a}uchli}},\ and\ \bibinfo {author} {\bibfnamefont
  {A.}~\bibnamefont {Browaeys}},\ }\bibfield  {title} {\bibinfo {title}
  {Quantum simulation of 2d antiferromagnets with hundreds of rydberg atoms},\
  }\href {https://doi.org/10.1038/s41586-021-03585-1} {\bibfield  {journal}
  {\bibinfo  {journal} {Nature}\ }\textbf {\bibinfo {volume} {595}},\ \bibinfo
  {pages} {233} (\bibinfo {year} {2021})}\BibitemShut {NoStop}%
\bibitem [{\citenamefont {Wu}\ \emph {et~al.}(2021)\citenamefont {Wu},
  \citenamefont {Liang}, \citenamefont {Tian}, \citenamefont {Yang},
  \citenamefont {Chen}, \citenamefont {Liu}, \citenamefont {Tey},\ and\
  \citenamefont {You}}]{Wu2021}%
  \BibitemOpen
  \bibfield  {author} {\bibinfo {author} {\bibfnamefont {X.}~\bibnamefont
  {Wu}}, \bibinfo {author} {\bibfnamefont {X.}~\bibnamefont {Liang}}, \bibinfo
  {author} {\bibfnamefont {Y.}~\bibnamefont {Tian}}, \bibinfo {author}
  {\bibfnamefont {F.}~\bibnamefont {Yang}}, \bibinfo {author} {\bibfnamefont
  {C.}~\bibnamefont {Chen}}, \bibinfo {author} {\bibfnamefont {Y.-C.}\
  \bibnamefont {Liu}}, \bibinfo {author} {\bibfnamefont {M.~K.}\ \bibnamefont
  {Tey}},\ and\ \bibinfo {author} {\bibfnamefont {L.}~\bibnamefont {You}},\
  }\bibfield  {title} {\bibinfo {title} {A concise review of rydberg atom based
  quantum computation and quantum simulation*},\ }\href
  {https://doi.org/10.1088/1674-1056/abd76f} {\bibfield  {journal} {\bibinfo
  {journal} {Chinese Physics B}\ }\textbf {\bibinfo {volume} {30}},\ \bibinfo
  {pages} {020305} (\bibinfo {year} {2021})}\BibitemShut {NoStop}%
\bibitem [{\citenamefont {Gallagher}(1994)}]{gallagher1994}%
  \BibitemOpen
  \bibfield  {author} {\bibinfo {author} {\bibfnamefont {T.~F.}\ \bibnamefont
  {Gallagher}},\ }\href {https://doi.org/10.1017/CBO9780511524530} {\emph
  {\bibinfo {title} {Rydberg Atoms}}},\ Cambridge Monographs on Atomic,
  Molecular and Chemical Physics\ (\bibinfo  {publisher} {Cambridge University
  Press},\ \bibinfo {year} {1994})\BibitemShut {NoStop}%
\bibitem [{\citenamefont {Safronova}\ \emph {et~al.}(2018)\citenamefont
  {Safronova}, \citenamefont {Porsev}, \citenamefont {Sanner},\ and\
  \citenamefont {Ye}}]{Safronova}%
  \BibitemOpen
  \bibfield  {author} {\bibinfo {author} {\bibfnamefont {M.~S.}\ \bibnamefont
  {Safronova}}, \bibinfo {author} {\bibfnamefont {S.~G.}\ \bibnamefont
  {Porsev}}, \bibinfo {author} {\bibfnamefont {C.}~\bibnamefont {Sanner}},\
  and\ \bibinfo {author} {\bibfnamefont {J.}~\bibnamefont {Ye}},\ }\bibfield
  {title} {\bibinfo {title} {Two clock transitions in neutral yb for the
  highest sensitivity to variations of the fine-structure constant},\ }\href
  {https://doi.org/10.1103/PhysRevLett.120.173001} {\bibfield  {journal}
  {\bibinfo  {journal} {Phys. Rev. Lett.}\ }\textbf {\bibinfo {volume} {120}},\
  \bibinfo {pages} {173001} (\bibinfo {year} {2018})}\BibitemShut {NoStop}%
\bibitem [{\citenamefont {Cheung}\ \emph {et~al.}(2020)\citenamefont {Cheung},
  \citenamefont {Safronova}, \citenamefont {Porsev}, \citenamefont {Kozlov},
  \citenamefont {Tupitsyn},\ and\ \citenamefont {Bondarev}}]{Cheung}%
  \BibitemOpen
  \bibfield  {author} {\bibinfo {author} {\bibfnamefont {C.}~\bibnamefont
  {Cheung}}, \bibinfo {author} {\bibfnamefont {M.~S.}\ \bibnamefont
  {Safronova}}, \bibinfo {author} {\bibfnamefont {S.~G.}\ \bibnamefont
  {Porsev}}, \bibinfo {author} {\bibfnamefont {M.~G.}\ \bibnamefont {Kozlov}},
  \bibinfo {author} {\bibfnamefont {I.~I.}\ \bibnamefont {Tupitsyn}},\ and\
  \bibinfo {author} {\bibfnamefont {A.~I.}\ \bibnamefont {Bondarev}},\
  }\bibfield  {title} {\bibinfo {title} {Accurate prediction of clock
  transitions in a highly charged ion with complex electronic structure},\
  }\href {https://doi.org/10.1103/PhysRevLett.124.163001} {\bibfield  {journal}
  {\bibinfo  {journal} {Phys. Rev. Lett.}\ }\textbf {\bibinfo {volume} {124}},\
  \bibinfo {pages} {163001} (\bibinfo {year} {2020})}\BibitemShut {NoStop}%
\bibitem [{\citenamefont {Kaur}\ \emph {et~al.}(2020)\citenamefont {Kaur},
  \citenamefont {Nakra}, \citenamefont {Arora}, \citenamefont {Li},\ and\
  \citenamefont {Sahoo}}]{Kaur}%
  \BibitemOpen
  \bibfield  {author} {\bibinfo {author} {\bibfnamefont {M.}~\bibnamefont
  {Kaur}}, \bibinfo {author} {\bibfnamefont {R.}~\bibnamefont {Nakra}},
  \bibinfo {author} {\bibfnamefont {B.}~\bibnamefont {Arora}}, \bibinfo
  {author} {\bibfnamefont {C.-B.}\ \bibnamefont {Li}},\ and\ \bibinfo {author}
  {\bibfnamefont {B.~K.}\ \bibnamefont {Sahoo}},\ }\bibfield  {title} {\bibinfo
  {title} {Accurate determination of energy levels, hyperfine structure
  constants, lifetimes and dipole polarizabilities of triply ionized tin
  isotopes},\ }\href {https://doi.org/10.1088/1361-6455/ab68b7} {\bibfield
  {journal} {\bibinfo  {journal} {Journal of Physics B: Atomic, Molecular and
  Optical Physics}\ }\textbf {\bibinfo {volume} {53}},\ \bibinfo {pages}
  {065002} (\bibinfo {year} {2020})}\BibitemShut {NoStop}%
\bibitem [{\citenamefont {Beloy}\ and\ \citenamefont
  {Derevianko}(2008)}]{Beloy}%
  \BibitemOpen
  \bibfield  {author} {\bibinfo {author} {\bibfnamefont {K.}~\bibnamefont
  {Beloy}}\ and\ \bibinfo {author} {\bibfnamefont {A.}~\bibnamefont
  {Derevianko}},\ }\bibfield  {title} {\bibinfo {title} {Second-order effects
  on the hyperfine structure of $p$ states of alkali-metal atoms},\ }\href
  {https://doi.org/10.1103/PhysRevA.78.032519} {\bibfield  {journal} {\bibinfo
  {journal} {Phys. Rev. A}\ }\textbf {\bibinfo {volume} {78}},\ \bibinfo
  {pages} {032519} (\bibinfo {year} {2008})}\BibitemShut {NoStop}%
\bibitem [{\citenamefont {Beloy}\ \emph {et~al.}(2006)\citenamefont {Beloy},
  \citenamefont {Safronova},\ and\ \citenamefont {Derevianko}}]{Derevianko}%
  \BibitemOpen
  \bibfield  {author} {\bibinfo {author} {\bibfnamefont {K.}~\bibnamefont
  {Beloy}}, \bibinfo {author} {\bibfnamefont {U.~I.}\ \bibnamefont
  {Safronova}},\ and\ \bibinfo {author} {\bibfnamefont {A.}~\bibnamefont
  {Derevianko}},\ }\bibfield  {title} {\bibinfo {title} {High-accuracy
  calculation of the blackbody radiation shift in the $^{133}\mathrm{Cs}$
  primary frequency standard},\ }\href
  {https://doi.org/10.1103/PhysRevLett.97.040801} {\bibfield  {journal}
  {\bibinfo  {journal} {Phys. Rev. Lett.}\ }\textbf {\bibinfo {volume} {97}},\
  \bibinfo {pages} {040801} (\bibinfo {year} {2006})}\BibitemShut {NoStop}%
\bibitem [{\citenamefont {Kratz}(1949)}]{Kratz}%
  \BibitemOpen
  \bibfield  {author} {\bibinfo {author} {\bibfnamefont {H.~R.}\ \bibnamefont
  {Kratz}},\ }\bibfield  {title} {\bibinfo {title} {The principal series of
  potassium, rubidium, and cesium in absorption},\ }\href
  {https://doi.org/10.1103/PhysRev.75.1844} {\bibfield  {journal} {\bibinfo
  {journal} {Phys. Rev.}\ }\textbf {\bibinfo {volume} {75}},\ \bibinfo {pages}
  {1844} (\bibinfo {year} {1949})}\BibitemShut {NoStop}%
\bibitem [{\citenamefont {McNally}\ \emph {et~al.}(1949)\citenamefont
  {McNally}, \citenamefont {Molnar}, \citenamefont {Hitchcock},\ and\
  \citenamefont {Oliver}}]{McNally}%
  \BibitemOpen
  \bibfield  {author} {\bibinfo {author} {\bibfnamefont {J.~R.}\ \bibnamefont
  {McNally}}, \bibinfo {author} {\bibfnamefont {J.~P.}\ \bibnamefont {Molnar}},
  \bibinfo {author} {\bibfnamefont {W.~J.}\ \bibnamefont {Hitchcock}},\ and\
  \bibinfo {author} {\bibfnamefont {N.~F.}\ \bibnamefont {Oliver}},\ }\bibfield
   {title} {\bibinfo {title} {High members of the principal series in
  caesium},\ }\href {https://doi.org/10.1364/JOSA.39.000057} {\bibfield
  {journal} {\bibinfo  {journal} {J. Opt. Soc. Am.}\ }\textbf {\bibinfo
  {volume} {39}},\ \bibinfo {pages} {57} (\bibinfo {year} {1949})}\BibitemShut
  {NoStop}%
\bibitem [{\citenamefont {Kleiman}(1962)}]{Kleiman}%
  \BibitemOpen
  \bibfield  {author} {\bibinfo {author} {\bibfnamefont {H.}~\bibnamefont
  {Kleiman}},\ }\bibfield  {title} {\bibinfo {title} {Interferometric
  measurements of cesium i},\ }\href {https://doi.org/10.1364/JOSA.52.000441}
  {\bibfield  {journal} {\bibinfo  {journal} {J. Opt. Soc. Am.}\ }\textbf
  {\bibinfo {volume} {52}},\ \bibinfo {pages} {441} (\bibinfo {year}
  {1962})}\BibitemShut {NoStop}%
\bibitem [{\citenamefont {Eriksson}\ and\ \citenamefont
  {Wen{\aa}ker}(1970)}]{Eriksson}%
  \BibitemOpen
  \bibfield  {author} {\bibinfo {author} {\bibfnamefont {K.~B.~S.}\
  \bibnamefont {Eriksson}}\ and\ \bibinfo {author} {\bibfnamefont
  {I.}~\bibnamefont {Wen{\aa}ker}},\ }\bibfield  {title} {\bibinfo {title} {New
  wavelength measurements in cs i},\ }\href
  {https://doi.org/10.1088/0031-8949/1/1/003} {\bibfield  {journal} {\bibinfo
  {journal} {Physica Scripta}\ }\textbf {\bibinfo {volume} {1}},\ \bibinfo
  {pages} {21} (\bibinfo {year} {1970})}\BibitemShut {NoStop}%
\bibitem [{\citenamefont {Sansonetti}\ \emph {et~al.}(1981)\citenamefont
  {Sansonetti}, \citenamefont {Andrew},\ and\ \citenamefont
  {Verges}}]{Sansonetti}%
  \BibitemOpen
  \bibfield  {author} {\bibinfo {author} {\bibfnamefont {C.~J.}\ \bibnamefont
  {Sansonetti}}, \bibinfo {author} {\bibfnamefont {K.~L.}\ \bibnamefont
  {Andrew}},\ and\ \bibinfo {author} {\bibfnamefont {J.}~\bibnamefont
  {Verges}},\ }\bibfield  {title} {\bibinfo {title} {Polarization, penetration,
  and exchange effects in the hydrogenlike $nf$ and $ng$ terms of cesium},\
  }\href {https://doi.org/10.1364/JOSA.71.000423} {\bibfield  {journal}
  {\bibinfo  {journal} {J. Opt. Soc. Am.}\ }\textbf {\bibinfo {volume} {71}},\
  \bibinfo {pages} {423} (\bibinfo {year} {1981})}\BibitemShut {NoStop}%
\bibitem [{\citenamefont {Lorenzen}\ \emph {et~al.}(1980)\citenamefont
  {Lorenzen}, \citenamefont {Weber},\ and\ \citenamefont {Niemax}}]{Lorenzen}%
  \BibitemOpen
  \bibfield  {author} {\bibinfo {author} {\bibfnamefont {C.~J.}\ \bibnamefont
  {Lorenzen}}, \bibinfo {author} {\bibfnamefont {K.~H.}\ \bibnamefont
  {Weber}},\ and\ \bibinfo {author} {\bibfnamefont {K.}~\bibnamefont
  {Niemax}},\ }\bibfield  {title} {\bibinfo {title} {Energies of the
  $n^2s_{12}$ and $n^2d_{32,52}$ states of cs},\ }\href
  {https://doi.org/https://doi.org/10.1016/0030-4018(80)90242-4} {\bibfield
  {journal} {\bibinfo  {journal} {Optics Communications}\ }\textbf {\bibinfo
  {volume} {33}},\ \bibinfo {pages} {271} (\bibinfo {year} {1980})}\BibitemShut
  {NoStop}%
\bibitem [{\citenamefont {Lorenzen}\ and\ \citenamefont
  {Niemax}(1983)}]{Lorenzen1983}%
  \BibitemOpen
  \bibfield  {author} {\bibinfo {author} {\bibfnamefont {C.~J.}\ \bibnamefont
  {Lorenzen}}\ and\ \bibinfo {author} {\bibfnamefont {K.}~\bibnamefont
  {Niemax}},\ }\bibfield  {title} {\bibinfo {title} {Non-monotonic variation of
  the quantum defect in cs $nd_{J}$ term series},\ }\href
  {https://doi.org/https://doi.org/10.1007/BF01415112} {\bibfield  {journal}
  {\bibinfo  {journal} {Z Physik A}\ }\textbf {\bibinfo {volume} {311}},\
  \bibinfo {pages} {249–250} (\bibinfo {year} {1983})}\BibitemShut {NoStop}%
\bibitem [{\citenamefont {Lorenzen}\ and\ \citenamefont
  {Niemax}(1984)}]{Lorenzen1984}%
  \BibitemOpen
  \bibfield  {author} {\bibinfo {author} {\bibfnamefont {C.~J.}\ \bibnamefont
  {Lorenzen}}\ and\ \bibinfo {author} {\bibfnamefont {K.}~\bibnamefont
  {Niemax}},\ }\bibfield  {title} {\bibinfo {title} {Precise quantum defects of
  $ns$, $np$ and $nd$ levels in cs i},\ }\href
  {https://doi.org/https://doi.org/10.1007/BF01419370} {\bibfield  {journal}
  {\bibinfo  {journal} {Z Physik A}\ }\textbf {\bibinfo {volume} {315}},\
  \bibinfo {pages} {127–133} (\bibinfo {year} {1984})}\BibitemShut {NoStop}%
\bibitem [{\citenamefont {O'Sullivan}\ and\ \citenamefont
  {Stoicheff}(1983)}]{Sullivan}%
  \BibitemOpen
  \bibfield  {author} {\bibinfo {author} {\bibfnamefont {M.~S.}\ \bibnamefont
  {O'Sullivan}}\ and\ \bibinfo {author} {\bibfnamefont {B.~P.}\ \bibnamefont
  {Stoicheff}},\ }\bibfield  {title} {\bibinfo {title} {Doppler-free two-photon
  absorption spectrum of cesium},\ }\href {https://doi.org/10.1139/p83-116}
  {\bibfield  {journal} {\bibinfo  {journal} {Canadian Journal of Physics}\
  }\textbf {\bibinfo {volume} {61}},\ \bibinfo {pages} {940} (\bibinfo {year}
  {1983})},\ \Eprint {https://arxiv.org/abs/https://doi.org/10.1139/p83-116}
  {https://doi.org/10.1139/p83-116} \BibitemShut {NoStop}%
\bibitem [{\citenamefont {Bjorkholm}\ and\ \citenamefont
  {Liao}(1976)}]{Bjorkholm}%
  \BibitemOpen
  \bibfield  {author} {\bibinfo {author} {\bibfnamefont {J.~E.}\ \bibnamefont
  {Bjorkholm}}\ and\ \bibinfo {author} {\bibfnamefont {P.~F.}\ \bibnamefont
  {Liao}},\ }\bibfield  {title} {\bibinfo {title} {Line shape and strength of
  two-photon absorption in an atomic vapor with a resonant or nearly resonant
  intermediate state},\ }\href {https://doi.org/10.1103/PhysRevA.14.751}
  {\bibfield  {journal} {\bibinfo  {journal} {Phys. Rev. A}\ }\textbf {\bibinfo
  {volume} {14}},\ \bibinfo {pages} {751} (\bibinfo {year} {1976})}\BibitemShut
  {NoStop}%
\bibitem [{\citenamefont {Liao}\ and\ \citenamefont {Bjorkholm}(1976)}]{Liao}%
  \BibitemOpen
  \bibfield  {author} {\bibinfo {author} {\bibfnamefont {P.~F.}\ \bibnamefont
  {Liao}}\ and\ \bibinfo {author} {\bibfnamefont {J.~E.}\ \bibnamefont
  {Bjorkholm}},\ }\bibfield  {title} {\bibinfo {title} {Measurement of the
  fine-structure splitting of the $4f$ state in atomic sodium using two-photon
  spectroscopy with a resonant intermediate state},\ }\href
  {https://doi.org/10.1103/PhysRevLett.36.1543} {\bibfield  {journal} {\bibinfo
   {journal} {Phys. Rev. Lett.}\ }\textbf {\bibinfo {volume} {36}},\ \bibinfo
  {pages} {1543} (\bibinfo {year} {1976})}\BibitemShut {NoStop}%
\bibitem [{\citenamefont {Goy}\ \emph {et~al.}(1982)\citenamefont {Goy},
  \citenamefont {Raimond}, \citenamefont {Vitrant},\ and\ \citenamefont
  {Haroche}}]{Goy}%
  \BibitemOpen
  \bibfield  {author} {\bibinfo {author} {\bibfnamefont {P.}~\bibnamefont
  {Goy}}, \bibinfo {author} {\bibfnamefont {J.~M.}\ \bibnamefont {Raimond}},
  \bibinfo {author} {\bibfnamefont {G.}~\bibnamefont {Vitrant}},\ and\ \bibinfo
  {author} {\bibfnamefont {S.}~\bibnamefont {Haroche}},\ }\bibfield  {title}
  {\bibinfo {title} {Millimeter-wave spectroscopy in cesium rydberg states.
  quantum defects, fine- and hyperfine-structure measurements},\ }\href
  {https://doi.org/10.1103/PhysRevA.26.2733} {\bibfield  {journal} {\bibinfo
  {journal} {Phys. Rev. A}\ }\textbf {\bibinfo {volume} {26}},\ \bibinfo
  {pages} {2733} (\bibinfo {year} {1982})}\BibitemShut {NoStop}%
\bibitem [{\citenamefont {Weber}\ and\ \citenamefont
  {Sansonetti}(1987)}]{Weber}%
  \BibitemOpen
  \bibfield  {author} {\bibinfo {author} {\bibfnamefont {K.~H.}\ \bibnamefont
  {Weber}}\ and\ \bibinfo {author} {\bibfnamefont {C.~J.}\ \bibnamefont
  {Sansonetti}},\ }\bibfield  {title} {\bibinfo {title} {Accurate energies of
  $ns$, $np$, $nd$, $nf$, and $ng$ levels of neutral cesium},\ }\href
  {https://doi.org/10.1103/PhysRevA.35.4650} {\bibfield  {journal} {\bibinfo
  {journal} {Phys. Rev. A}\ }\textbf {\bibinfo {volume} {35}},\ \bibinfo
  {pages} {4650} (\bibinfo {year} {1987})}\BibitemShut {NoStop}%
\bibitem [{\citenamefont {Moore}\ \emph {et~al.}(2020)\citenamefont {Moore},
  \citenamefont {Duspayev}, \citenamefont {Cardman},\ and\ \citenamefont
  {Raithel}}]{Moore}%
  \BibitemOpen
  \bibfield  {author} {\bibinfo {author} {\bibfnamefont {K.}~\bibnamefont
  {Moore}}, \bibinfo {author} {\bibfnamefont {A.}~\bibnamefont {Duspayev}},
  \bibinfo {author} {\bibfnamefont {R.}~\bibnamefont {Cardman}},\ and\ \bibinfo
  {author} {\bibfnamefont {G.}~\bibnamefont {Raithel}},\ }\bibfield  {title}
  {\bibinfo {title} {Measurement of the rb $g$-series quantum defect using
  two-photon microwave spectroscopy},\ }\href
  {https://doi.org/10.1103/PhysRevA.102.062817} {\bibfield  {journal} {\bibinfo
   {journal} {Phys. Rev. A}\ }\textbf {\bibinfo {volume} {102}},\ \bibinfo
  {pages} {062817} (\bibinfo {year} {2020})}\BibitemShut {NoStop}%
\bibitem [{\citenamefont {Bai}\ \emph {et~al.}(2020)\citenamefont {Bai},
  \citenamefont {Bai}, \citenamefont {Han}, \citenamefont {Jiao}, \citenamefont
  {Zhao},\ and\ \citenamefont {Jia}}]{Bai}%
  \BibitemOpen
  \bibfield  {author} {\bibinfo {author} {\bibfnamefont {J.}~\bibnamefont
  {Bai}}, \bibinfo {author} {\bibfnamefont {S.}~\bibnamefont {Bai}}, \bibinfo
  {author} {\bibfnamefont {X.}~\bibnamefont {Han}}, \bibinfo {author}
  {\bibfnamefont {Y.}~\bibnamefont {Jiao}}, \bibinfo {author} {\bibfnamefont
  {J.}~\bibnamefont {Zhao}},\ and\ \bibinfo {author} {\bibfnamefont
  {S.}~\bibnamefont {Jia}},\ }\bibfield  {title} {\bibinfo {title} {Precise
  measurements of polarizabilities of cesium $ns$ rydberg states in an
  ultra-cold atomic ensemble},\ }\href
  {https://doi.org/10.1088/1367-2630/abaf30} {\bibfield  {journal} {\bibinfo
  {journal} {New Journal of Physics}\ }\textbf {\bibinfo {volume} {22}},\
  \bibinfo {pages} {093032} (\bibinfo {year} {2020})}\BibitemShut {NoStop}%
\bibitem [{\citenamefont {Reinhard}\ \emph {et~al.}(2007)\citenamefont
  {Reinhard}, \citenamefont {Liebisch}, \citenamefont {Knuffman},\ and\
  \citenamefont {Raithel}}]{Reinhard}%
  \BibitemOpen
  \bibfield  {author} {\bibinfo {author} {\bibfnamefont {A.}~\bibnamefont
  {Reinhard}}, \bibinfo {author} {\bibfnamefont {T.~C.}\ \bibnamefont
  {Liebisch}}, \bibinfo {author} {\bibfnamefont {B.}~\bibnamefont {Knuffman}},\
  and\ \bibinfo {author} {\bibfnamefont {G.}~\bibnamefont {Raithel}},\
  }\bibfield  {title} {\bibinfo {title} {Level shifts of rubidium rydberg
  states due to binary interactions},\ }\href
  {https://doi.org/10.1103/PhysRevA.75.032712} {\bibfield  {journal} {\bibinfo
  {journal} {Phys. Rev. A}\ }\textbf {\bibinfo {volume} {75}},\ \bibinfo
  {pages} {032712} (\bibinfo {year} {2007})}\BibitemShut {NoStop}%
\bibitem [{\citenamefont {Han}\ \emph {et~al.}(2018)\citenamefont {Han},
  \citenamefont {Bai}, \citenamefont {Jiao}, \citenamefont {Hao}, \citenamefont
  {Xue}, \citenamefont {Zhao}, \citenamefont {Jia},\ and\ \citenamefont
  {Raithel}}]{Han2018}%
  \BibitemOpen
  \bibfield  {author} {\bibinfo {author} {\bibfnamefont {X.}~\bibnamefont
  {Han}}, \bibinfo {author} {\bibfnamefont {S.}~\bibnamefont {Bai}}, \bibinfo
  {author} {\bibfnamefont {Y.}~\bibnamefont {Jiao}}, \bibinfo {author}
  {\bibfnamefont {L.}~\bibnamefont {Hao}}, \bibinfo {author} {\bibfnamefont
  {Y.}~\bibnamefont {Xue}}, \bibinfo {author} {\bibfnamefont {J.}~\bibnamefont
  {Zhao}}, \bibinfo {author} {\bibfnamefont {S.}~\bibnamefont {Jia}},\ and\
  \bibinfo {author} {\bibfnamefont {G.}~\bibnamefont {Raithel}},\ }\bibfield
  {title} {\bibinfo {title} {Cs $62{D}_{J}$ rydberg-atom macrodimers formed by
  long-range multipole interaction},\ }\href
  {https://doi.org/10.1103/PhysRevA.97.031403} {\bibfield  {journal} {\bibinfo
  {journal} {Phys. Rev. A}\ }\textbf {\bibinfo {volume} {97}},\ \bibinfo
  {pages} {031403} (\bibinfo {year} {2018})}\BibitemShut {NoStop}%
\bibitem [{\citenamefont {Han}\ \emph {et~al.}(2019)\citenamefont {Han},
  \citenamefont {Bai}, \citenamefont {Jiao}, \citenamefont {Raithel},
  \citenamefont {Zhao},\ and\ \citenamefont {Jia}}]{Han2019}%
  \BibitemOpen
  \bibfield  {author} {\bibinfo {author} {\bibfnamefont {X.}~\bibnamefont
  {Han}}, \bibinfo {author} {\bibfnamefont {S.}~\bibnamefont {Bai}}, \bibinfo
  {author} {\bibfnamefont {Y.}~\bibnamefont {Jiao}}, \bibinfo {author}
  {\bibfnamefont {G.}~\bibnamefont {Raithel}}, \bibinfo {author} {\bibfnamefont
  {J.}~\bibnamefont {Zhao}},\ and\ \bibinfo {author} {\bibfnamefont
  {S.}~\bibnamefont {Jia}},\ }\bibfield  {title} {\bibinfo {title} {Adiabatic
  potentials of cesium $(nd_j)^2$ rydberg–rydberg macrodimers},\ }\href
  {https://doi.org/10.1088/1361-6455/ab1371} {\bibfield  {journal} {\bibinfo
  {journal} {J. Phys B}\ }\textbf {\bibinfo {volume} {52}},\ \bibinfo {pages}
  {135102} (\bibinfo {year} {2019})}\BibitemShut {NoStop}%
\bibitem [{\citenamefont {Li}\ \emph {et~al.}(2003)\citenamefont {Li},
  \citenamefont {Mourachko}, \citenamefont {Noel},\ and\ \citenamefont
  {Gallagher}}]{Li}%
  \BibitemOpen
  \bibfield  {author} {\bibinfo {author} {\bibfnamefont {W.}~\bibnamefont
  {Li}}, \bibinfo {author} {\bibfnamefont {I.}~\bibnamefont {Mourachko}},
  \bibinfo {author} {\bibfnamefont {M.~W.}\ \bibnamefont {Noel}},\ and\
  \bibinfo {author} {\bibfnamefont {T.~F.}\ \bibnamefont {Gallagher}},\
  }\bibfield  {title} {\bibinfo {title} {Millimeter-wave spectroscopy of cold
  rb rydberg atoms in a magneto-optical trap: Quantum defects of the $ns$,
  $np$, and $nd$ series},\ }\href {https://doi.org/10.1103/PhysRevA.67.052502}
  {\bibfield  {journal} {\bibinfo  {journal} {Phys. Rev. A}\ }\textbf {\bibinfo
  {volume} {67}},\ \bibinfo {pages} {052502} (\bibinfo {year}
  {2003})}\BibitemShut {NoStop}%
\bibitem [{\citenamefont {Fredriksson}\ \emph {et~al.}(1980)\citenamefont
  {Fredriksson}, \citenamefont {Lundberg},\ and\ \citenamefont
  {Svanberg}}]{Fredriksson1980}%
  \BibitemOpen
  \bibfield  {author} {\bibinfo {author} {\bibfnamefont {K.}~\bibnamefont
  {Fredriksson}}, \bibinfo {author} {\bibfnamefont {H.}~\bibnamefont
  {Lundberg}},\ and\ \bibinfo {author} {\bibfnamefont {S.}~\bibnamefont
  {Svanberg}},\ }\bibfield  {title} {\bibinfo {title} {Fine- and
  hyperfine-structure investigation in the $5^{2}d\ensuremath{-}n^{2}f$ series
  of cesium},\ }\href {https://doi.org/10.1103/PhysRevA.21.241} {\bibfield
  {journal} {\bibinfo  {journal} {Phys. Rev. A}\ }\textbf {\bibinfo {volume}
  {21}},\ \bibinfo {pages} {241} (\bibinfo {year} {1980})}\BibitemShut
  {NoStop}%
\end{thebibliography}%

\end{document}